\documentclass[11pt,a4paper]{article}
\RequirePackage[english]{babel}
\RequirePackage[T1]{fontenc}
\RequirePackage{graphicx}
\usepackage{tikz}
\usepackage{tikz-3dplot}
\usetikzlibrary{3d}

\begin{document}

\title{\begin{flushright}
Nikhef 2015-015  
\end{flushright} 
\vspace{10mm}
The center of lateral iso-density contours for inclined cosmic air showers.}
\author{J.M.C. Montanus \\ \\ Nikhef, Science Park 105, 1098 XG, Amsterdam \\ \\ hansm@nikhef.nl}
\date{\today}
\maketitle

\begin{abstract}
The horizontal lateral density of a cosmic air shower with a non-zero zenith angle is asymmetric. The asymmetry consist of a stretching of the iso-density contours to ellipses and to a shift of the center of the elliptic contours with respect to the core of the shower. The shift is caused by atmospheric attenuation. The modeling of the attenuation results in an equation for the shift as a function of zenith angle and the size of the iso-density contours. A more accurate equation is obtained by investigating the shift in lateral densities of simulated showers. It is shown how the shift can be incorporated in an elliptic lateral density function. A linear approximation for the shift allows for an analytical solution for the shifted elliptic density. Its predictions for the polar variations of the density are compared with data of simulated showers.
\end{abstract}

\newpage
\section{Introduction}
\begin{figure}[htbp]
\begin{center}
\includegraphics[width=8cm]{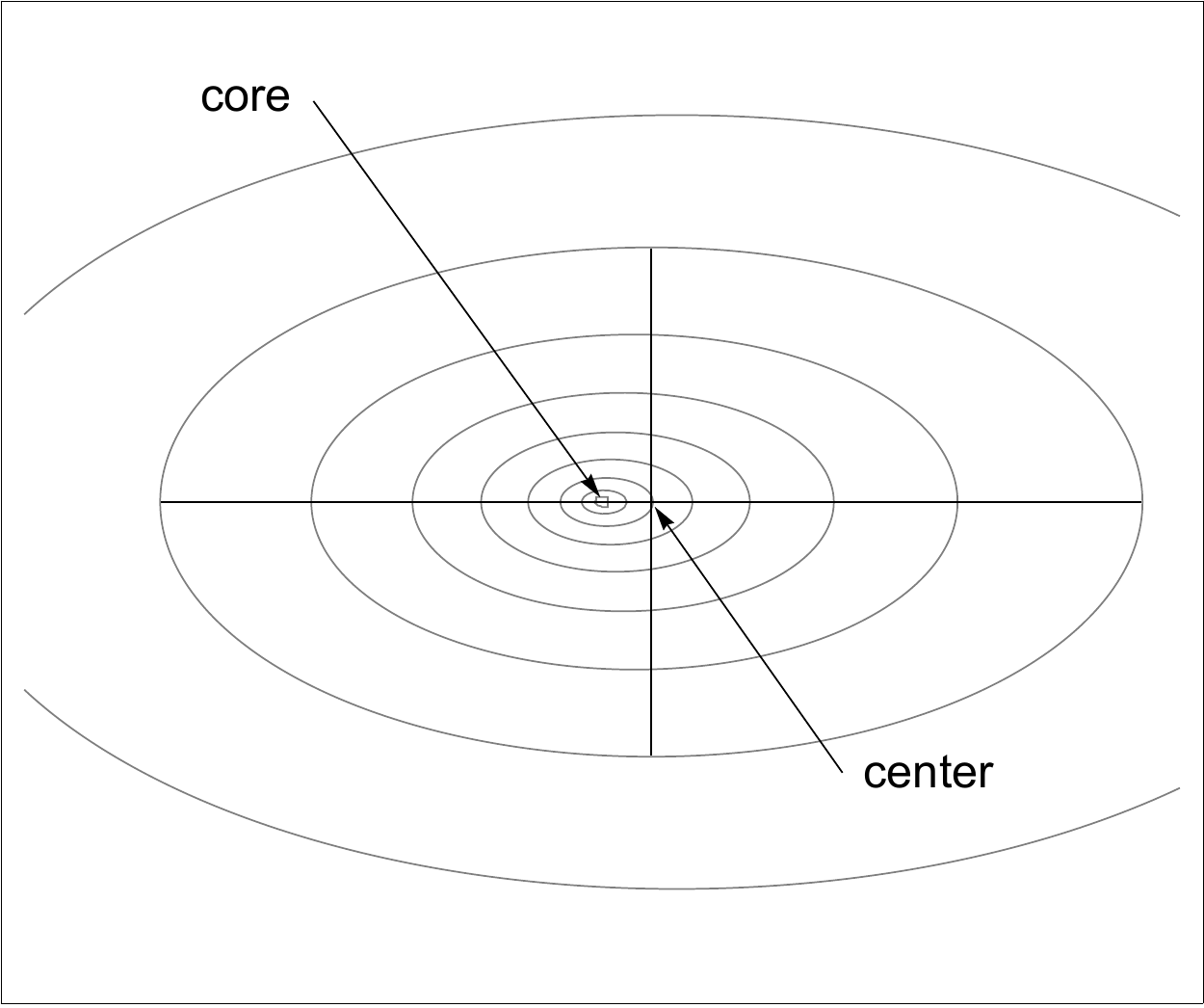}
\caption{Impression of the lateral density by means of iso-density contours and the different positions of the shower core and the point of intersection of the major axis and the minor axis, the center, of an iso-density contour.}
\label{corecentervis}
\end{center}
\end{figure}

\begin{figure}[htbp]
\begin{center}
\includegraphics[width=7cm]{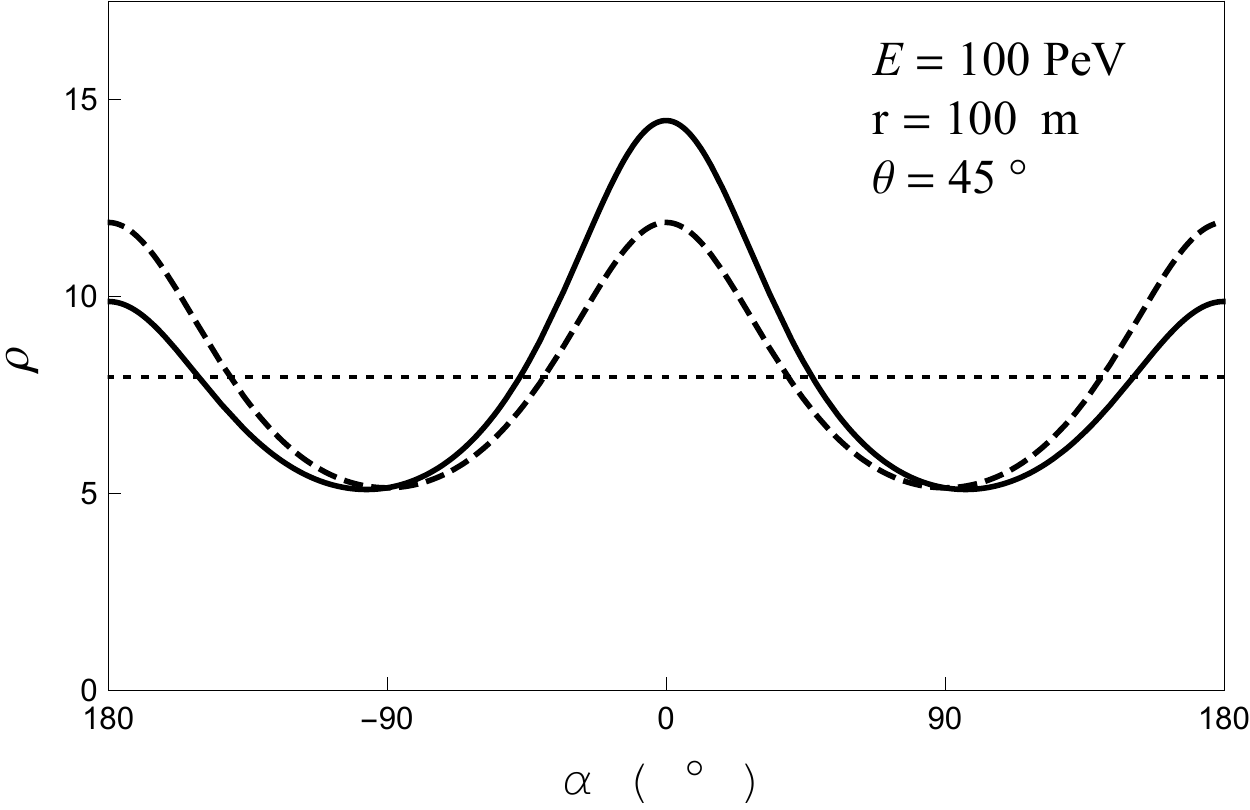}
\caption{The polar density according to projection with a shift (solid) and projection without a shift (dashed) for an arbitrary 100 PeV shower with zenith angle $45^\circ$ at a distance 100 m from the core. The horizontal line (dotted) is the mean density.}
\label{shiftellcomp}
\end{center}
\end{figure}

A lateral density function (LDF) describes the density as a function of the radius with respect to the core of a shower. For vertical air showers the horizontal plane coincides with the plane of the front of the shower and the iso-density contours in the horizontal plane are circles. A polar symmetric LDF is of application for vertical showers and for polar averaged densities of inclined showers. For inclined showers the iso-density contours are rather ellipses \cite{Dova1999,Pryke}.  As known, the centers of the elliptic iso-density contours do not coincide with the shower core, see Fig. \ref{corecentervis}.
\\
The distance between the shower core and the center of an elliptic contour will be denoted as the `shift'. The application of an elliptic LDF instead of a polar symmetric LDF increases the accuracy of the reconstruction of an inclined air shower observed with detectors in a horizontal plane. The accuracy of the reconstruction of an inclined air shower can be increased further if the shift is taken into account. In order to give already an impression an LDF-A solely based on the projection, thus without a shift, and an LDF-A including the shift, are both plotted for the polar density for an average 100 PeV shower with zenith angle $45^\circ$ at a distance 100 m from the core in Fig. \ref{shiftellcomp}. We see the additional angular density variations caused by the shift are of the same order as the angular density variations caused by the projection. This suggest that if the ellipticity is taken into account for reconstruction purposes, then the shift might be taken into consideration as well.
\\
The main purpose of the paper is the construction of an asymmetric density function which includes the shift. To this end the shift will be investigated for different primary energies and different zenith angles. The shift is caused by the attenuation of the shower. For the electron part this is the atmospheric attenuation which can be modeled to a certain extent. For the muon part the attenuation is mainly due to decay. The decay of the muons can be modeled. However, a decaying muon contributes an electron to the electron part. The two distribution are therefore intertwined. It therefore does not make much sense to consider the shift for electrons and muons separately. Besides it would require to consider the ratio of the densities of the electrons and muons in a model. Although we are interested in the shift   of the combined density of electron and muons together, we will for the model restrict to atmospheric attenuation. Even with this restriction the accuracy of the model is limited for several reasons of which the ignorance of local attenuation is the most important. Nevertheless, the model result gives an indication of the way the shift depends on zenith angle and on distance to the core. Accurate values for the shift are determined from lateral densities of MC showers. In plots of the determined shifts the model prediction will still be plotted for reasons of comparison. Furthermore it will be shown how the asymmetric LDF including the shift is constructed from a polar symmetric LDF. To avoid length we will denote a polar symmetric LDF just as LDF and an asymmetric LDF as LDF-A. For a clear distinction we will denote the density in the front plane as $\rho$ and the asymmetric density in the horizontal plane as $\nu$.
\\ \\
The contents of the paper can be divided in three parts: the modeling of the shift (Section 2 - 5), the determination of the shift (Section 6 - 8) and the construction of a polar density function including the shift (Section 9 - 10). In Section 2 we consider a cylinder model for the shower in a suitable coordinate system. The consequences of the cylindrical projection will be considered for the situation without and with a shift. In Section 3 we model the effect of atmospheric attenuation of the shower on the lateral density. An analytical approximation for the shift is derived in Section 4. In Section 5 we obtain a comparable result on the basis of a cone model for the shower. In Section 6 a description is given of the method used for the investigation of the shift on the basis of horizontal densities of simulated showers. In Section 7 some general results will be presented as obtained from simulated showers. In Section 8 we focus on the behavior of the shift for the combined density of electrons and muons together. The muon energy deposit in scintillators is practically similar to the electron energy deposit \cite{Sima2011}. A combined lateral density of electrons and muons is therefore of interest for scintillator based observatories. The detecting efficiency of scintillator detectors become small for densities below 0.5 m$^{-2}$. We therefore will focus on combined densities larger than 0.5 m$^{-2}$. We will see that in this region the shift is independent of shower size. Moreover, the relation between shift and the size of the elliptic contour is almost linear. The shift of the combined density will be compared with a proposed linear approximation. In Section 9 the polar density is considered. It is shown how to convert an LDF to the LDF-A including the shift as a function of radius $r$ and polar angle $\alpha$ and parameterized by the zenith angle $\theta$. For the outline of the procedure we conveniently restrict to a proposed linear approximation for the shift since it allows for an analytical solution for the LDF-A. To illustrate the procedure the LDF-A will be constructed explicitly for example LDF's of three simulated showers in Section 10. The predictions of the constructed LDF-A will be compared with the polar density of the simulated showers. In Section 11 the paper is concluded with a brief summary.

\section{Cylinder model}
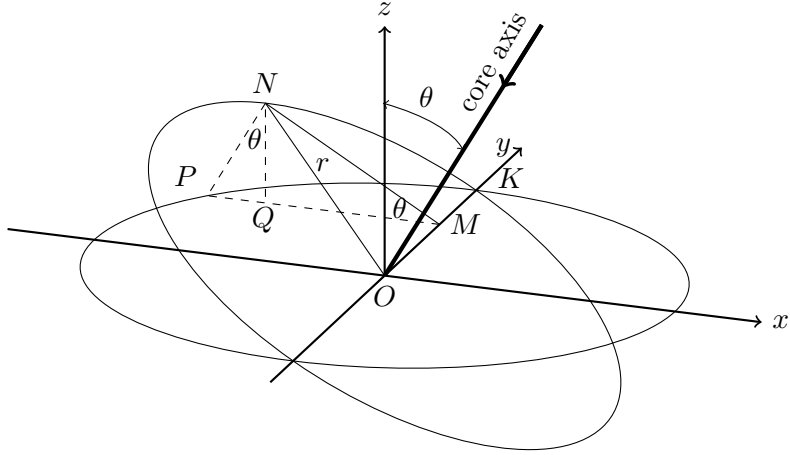
\begin{figure}
\begin{center}
\tdplotsetmaincoords{70}{20}
\begin{tikzpicture}[scale=4.4,tdplot_main_coords]
\tikzstyle{grid}=[thin,color=red,tdplot_rotated_coords]

\draw[thick,->] (0,0,0) -- (1.2,0,0) node[anchor=west]{$x$};
\draw[thick] (0,0,0) -- (-1.2,0,0) ;
\draw[thick,->] (0,0,0) -- (0,1.2,0) node[anchor=east]{$y$};
\draw[thick] (0,0,0) -- (0,-1,0);
\draw[thick,->] (0,0,0) -- (0,0,.8) node[anchor=south]{$z$};
\draw (0,0,0) circle [x radius=.924,y radius=.8];
\draw[dashed] (0,.48,0) -- (-.739,.48,0) node[anchor=south east]{$P$};
\draw[ultra thick] (0,.0,0) -- (.5,0,.866);
\draw[ultra thick,->] (.5,0,.866) -- (.375,0,.6495);
\draw[dashed] (-.554,.48,.32) -- (-.739,.48,0);
\draw[thin] (0,-.1,0) -- (0,0,0) node[anchor= north]{$O$};;
\draw[dashed] (-.554,.48,.30) -- (-.554,.48,0);
\draw[thin] (0,0,0) -- (0,.48,0) node[anchor=west]{$M$};
\node[rotate=60] at (.34,0,.73) {core axis};
\draw[<->] (.25,0,.433) arc (30:70:0.7) ;
\node[] at (-.153,.553,0.01)  {$\theta$};
\node[] at (-.14,.75,.30)  {$\theta$};
\node[] at (-.588,.48,.2)  {$\theta$};
\node[] at (-.51,.34,0)  {$Q$};
\node[] at (.06,.94,0)  {$K$};
\node[] at (-.32,.34,.2)  {$r$};

\tdplotsetrotatedcoords{0}{30}{0}
\draw[tdplot_rotated_coords] (0,0,0) circle [x radius=.8,y radius=.8];
\draw[tdplot_rotated_coords] (0,.48,0) -- (-.64,.48,0) node[anchor=south]{$N$};
\draw[tdplot_rotated_coords] (0,0,0) -- (-.64,.48,0);

\end{tikzpicture}
\caption{Front of inclined air shower falling on a horizontal surface.}
\label{contour1}
\end{center}
\end{figure}

Atmospheric attenuation has a large effect on inclined showers \cite{Dova2003}. One of the consequences is that it shifts the center of an elliptic iso-density contour. To model it we will first assume that all the particles run parallel with the shower core at the moment of arrival. Furthermore, we assume that contours of equal density are circles in the plane perpendicular to the shower direction. For the coordinate system we take the $x$ and $y$ axes in the horizontal plane and the $z$ axis in the upward vertical direction. The origin is taken at the position where the shower core axis intersects the horizontal plane. The azimuth angle of the shower is, anti-clockwise, with respect to the positive $x$-axis. Without loss of generality we consider inclined showers with zero azimuthal angle, thus with the shower core in the $x$,$z$-plane. The situation is schematically shown in Fig. \ref{contour1}. The tilted circle is perpendicular to the shower direction. At the moment the core reaches the surface in the origin of the coordinate system the shower front intersects the horizontal plane at the $y$-axis. We take a point $N$ on the tilted circle. Its distance with respect to the origin is $r$. If the direction of the shower particles is conveniently assumed parallel to the shower core, the projection of $N$ on the horizontal plane is point $P$. $M$ is a point on the $y$-axis with identical $y$-coordinate as $N$ and $P$. The angle between $NM$ and $PM$ and the angle between $NP$ and the vertical axis both are equal to the zenith angle $\theta$. From the geometry it follows
\begin{equation}\label{1}
MN^2=ON^2-OM^2=r^2-y^2 \ ,f
\end{equation}
\begin{equation}\label{2}
r^2=x^2 \cos^2 \theta +y^2 
\end{equation}  
and 
\begin{equation}\label{3}
NP^2=x^2 \sin^2 \theta  \ 
\end{equation} 
Here and in the sequel ($x$,$y$) denote the coordinates of $P$ in the horizontal plane. 
\\
\\
Without attenuation the asymmetry would be solely caused by the projection of the shower plane onto the horizontal observation plane. Alternatively, the intersection of a slant cylinder with a horizontal plane is an ellipse. The projection along the shower core axis means that the density along $NM$ is projected to the larger $PM$. As a consequence the density $\nu$ at the horizontal plane is smaller than the density $\rho$ of the inclined shower front by a factor $\cos \theta$:
\begin{equation}\label{4}
\nu(x,y)=\rho (r) \cos \theta  \ .
\end{equation} 
At the same time the iso-density contours are stretched to ellipses satisfying Eq. (\ref{2}). The horizontal ellipse and the inclined circle intersect each other and the positive $y$-axis at $K$. Denoting the $y$-coordinate of $K$ as $k$ we obtain 
\begin{equation}\label{5}
x^2 \cos^2 \theta +y^2 = k^2 \ .
\end{equation} 
This is an ellipse whose semi-major axis $a$ and semi-minor axis $b$ are related to each other via $b=a \cos \theta$ and where the center of the ellipse coincides with the shower core. 
\\ \\
Denoting the $x$-coordinate of the shifted center as $x_M$ the general equation for a shifted ellipse is
\begin{equation}\label{5a}
(x-x_M)^2 \cos^2 \theta +y^2 = b^2 \ ,
\end{equation} 
where $b$ is the size of the semi-minor axis. Since $y=k$ if $x=0$ we also obtain
\begin{equation}\label{5b}
x_M^2 \cos^2 \theta +k^2 = b^2 \ .
\end{equation} 
From the latter two equations we can write the equation of the shifted ellipse also as
\begin{equation}\label{5c}
(x^2-2x x_M) \cos^2 \theta +y^2 = k^2 \ .
\end{equation} 
The equation can be solved for $k$ after we have determined $x_M$ as a function of $k$ and $\theta$. By means of the solution the LDF-A can be constructed.

\section{Modeling attenuation}
At the early stages of the longitudinal development the size of a shower increases. After the shower size has reached a maximum it approximately falls of exponentially with atmospheric depth. The attenuation length $\lambda$ is about 185 g cm$^{-2}$ \cite{CiampaClay,Antoni2003}. A consequence of the attenuation of the shower during the traverse from $N$ to $P$ is that the density of shower particles is decreased by a factor $e^{-\Delta X / \lambda}$, where $\Delta X$ is the additional atmospheric depth met by shower particles between $N$ and $P$. The atmospheric depth exponentially decreases with altitude with a characteristic length of about 8 km. Except for shower with a very large energy and very large inclination the atmospheric depth between $N$ and $P$ approximately is constant.  At the surface of the earth the increase $\Delta X$ is approximately equal to 0.13 g cm$^{-2}$ for every meter travelled through the air. Hence,
\begin{equation}\label{6}
\nu(x,y)=\rho (r) \cdot  e^{- \xi \cdot NP} \cdot \cos \theta \ ,
\end{equation}  
where $\xi = 0.13 / 185 \approx 7.0 \cdot 10^{-4}$ m$^{-1}$.
With the substitution of the Eq. (\ref{3}) for $NP$ this is:
\begin{equation}\label{7}
\nu(x,y)=\rho (r)  \cdot  e^{\xi x \sin \theta} \cdot \cos \theta  \ ,
\end{equation} 
where, according to Eq. (\ref{2}),  $r=\sqrt{x^2 \cos^2 \theta +y^2}$. This is the basic equation for the analysis. It accounts for the attenuation at the late part of the inclined shower and for the reverse at the early part. As a consequence it leads to a shift of the elliptic density in the horizontal plane. The performance of the reconstruction of shower core positions should improve if the polar density function is modified for the shift. In our coordinate system the late part of the shower is at the negative $x$-axis. Notice that a negative value for $x$ leads to a decrease of the density. For $x=0$, $y=k$ we have 
\begin{equation}\label{8}
\nu(0,k)=\rho (k) \cdot   \cos \theta  \ .
\end{equation} 
To obtain the iso-density contour in the horizontal plane through $(0,k)$ we have to solve the equation $\nu (x,y) = \nu (0,k)$. That is, we have to solve the equation
\begin{equation}\label{8}
\nu(x,y)=\rho (k) \cdot   \cos \theta
\end{equation} 
or, more explicitly,
\begin{equation}\label{9}
\rho(r) \cdot e^{- \xi x \sin \theta}  = \rho(k) \ ,
\end{equation} 
where $r=\sqrt{x^2 \cos^2 \theta +y^2}$. In the next section we will derive analytically a first order solution for this equation. 

\section{Analytical approximation}
\begin{figure}[htbp]
\begin{center}
\includegraphics[width=8.5cm]{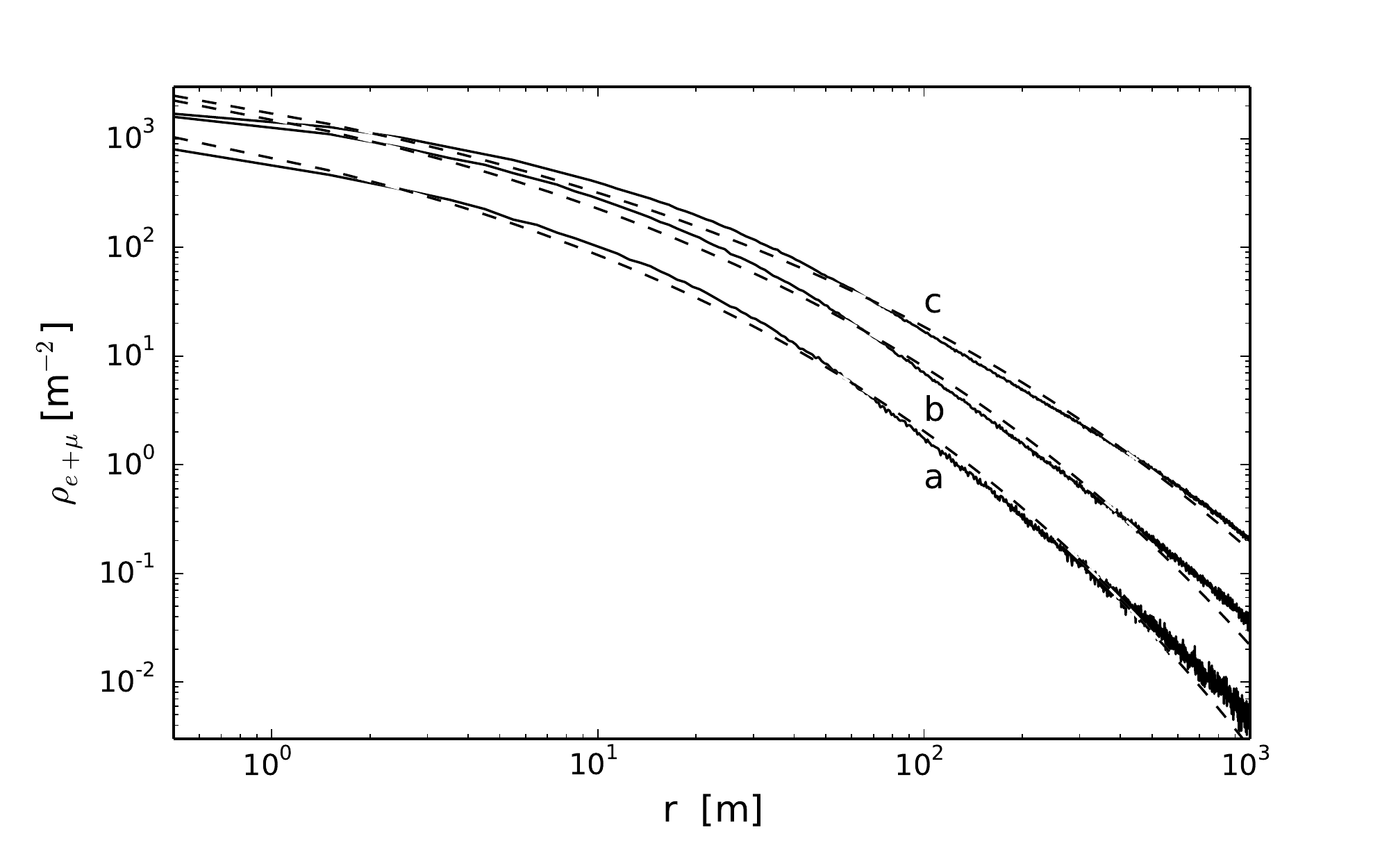}
\caption{Polar averaged lateral density of electrons and muons together for the three showers \textbf{a}, \textbf{b} and \textbf{c} as given in the text. The dashed curves are the approximations by the exponential function as given in the text.}
\label{vwkfit}
\end{center}
\end{figure}

The key in the following analysis is the observation that the lateral density can be roughly described by  the following exponential function:
\begin{equation}\label{10}
\rho(r) \propto  e^{-(r/r_0)^w} \ .
\end{equation} 
In Fig. \ref{vwkfit} the polar averaged combined density, binned with bin-width 1 m, and their approximation by the exponential function are plotted for three showers: one with energy $10^{16}$ and zenith angle $30^\circ$, one with energy $10^{17}$ and zenith angle $45^\circ$ and one with energy $10^{18}$ and zenith angle $52.5^\circ$ respectively denoted as shower \textbf{a}, \textbf{b} and \textbf{c}.
\\
It is evident that the approximation is not particularly good. The exponential function is solely intended as a toy function. The values for the parameters of shower  \textbf{a}, \textbf{b} and \textbf{c} respectively are 0.022, 0.026, and 0.017 meter for $r_0$ and 0.25, 0.25, and 0.22 for $w$. In the remainder of the analysis we will solely use $r_0 = 0.025$ m and $w = 0.25$.  By means of the `toy' function for the lateral density the equation  $\nu (x,y) = \nu (0,k)$ can be written as
\begin{equation}\label{11}
e^{-(r/r_0)^w} e^{ \xi x \sin \theta} =  e^{-(k/r_0)^w} \ .
\end{equation} 
Hence,
\begin{equation}\label{12}
r^2  = k^2 \left(1+ \frac{\xi x \sin \theta}{(k/r_0)^w} \right)^{2/w}  \ .
\end{equation} 
The latter can be expressed as 
\begin{equation}\label{13}
r^2  = k^2 \left(1+x w \eta \cos \theta \right)^{2/w}  \ ,
\end{equation} 
where
\begin{equation}\label{14}
\eta = \frac{\xi}{w(k/r_0)^w}  \tan \theta \ .
\end{equation} 
Since $xw\eta \cos \theta <1$ for $x$ smaller than $10^4$ m we will take a first order approximation for the right hand side of Eq. (\ref{13}): 
\begin{equation}\label{15}
r^2 \approx k^2 \left(1+2 x \eta \cos \theta  \right)  \ .
\end{equation} 
With the substitution of Eq. (\ref{2}) for $r$ and some rearrangement we obtain
\begin{equation}\label{16}
\left(x \cos \theta - k^2 \eta \right)^2 +y^2  \approx k^2 \left(1+k^2 \eta^2 \right) \ .
\end{equation}
From the comparison with Eqs. (\ref{5a}) and (\ref{5b}) we find
\begin{equation}\label{16a}
b^2 \approx k^2+k^4 \eta^2
\end{equation}
and
\begin{equation}\label{17}
x_M \approx \frac{k^2 \eta}{\cos \theta} \ .
\end{equation}
With the substitution of the Eq. (\ref{14}) for $\eta$ we obtain the following model prediction for the shift:
\begin{equation}\label{18}
x_M = \frac{\xi \cdot r_0^w}{w} \cdot k^{2-w} \cdot \frac{\tan \theta}{\cos \theta} \ .
\end{equation}
Since $x_M >0$ the shower attenuation does shift the center of the ellipse towards the early part of the shower. The shower attenuation did not change the eccentricity, which is equal to $\sin \theta$. Substituting $\xi=7.0 \cdot 10^{-4}$ m$^{-1}$, $r_0 = 0.025$ m and $w = 0.25$, we obtain as the `cylinder' model prediction for the shift in m:
\begin{equation}\label{19}
x_M =1.1 \cdot 10^{-3} \cdot k^{1.75} \cdot \frac{\tan \theta}{\cos \theta} \ .
\end{equation}

\section{Cone model}
Another model is the cone model. That is, we consider paths from apex $A$ through the horizontal plane at ground level. The differences in experienced atmospheric depth, due to the different path lengths, will be translated in attenuation. To this end we consider a shower cone with apex $A$ in the same coordinate system as in Fig. \ref{contour1}. As before, we regard inclined showers with zero azimuthal angle, thus with the shower core in the $x$,$z$-plane. For positions inside the cone the opening angle is denoted as $\delta$. The situation is schematically shown in Fig. \ref{contour2}. 
\begin{figure}[htbp]
\begin{center}
\tdplotsetmaincoords{70}{20}
\begin{tikzpicture}[scale=4,tdplot_main_coords]
\tikzstyle{grid}=[thin,color=red,tdplot_rotated_coords]
\draw[thick,->] (0,0,0) -- (1.0,0,0) node[anchor=west]{$x$};
\draw[thick] (0,0,0) -- (-1.2,0,0) ;
\draw[thick,->] (0,0,0) -- (0,1.2,0) node[anchor=east]{$y$};
\draw[thick] (0,0,0) -- (0,-1,0);
\draw[thick,->] (0,0,0) -- (0,0,.95) node[anchor=south]{$z$};
\draw (-1.1,.0,0) arc (180:360:0.935 );
\draw (-1.1,.0,0) arc (180:77:0.805 );
\draw (.767,0,0) arc (10:87:0.95);
\draw[dashed] (0,0,0) -- (-.55,.48,0);
\draw[dashed] (-.55,.48,.32) -- (-.88,.56,0);
\draw[dashed] (-.55,.48,0) -- (-.88,.56,0) node[anchor=south ]{$P$};
\draw[ultra thick] (0,.0,0) -- (1.,0,1.732);
\draw[thin] (1,0,1.732) -- (-.55,.48,.32);
\draw[thin] (1,0,1.732) -- (.72,0,-.42);
\draw[thin] (0,0,0) -- (.72,0,-.42);
\draw[ultra thick,->] (.5,0,.866) -- (.375,0,.6495);
\draw[thin] (0,-.1,0) -- (0,0,0) node[anchor= north]{$O$};;
\draw[dashed] (-.554,.48,.30) -- (-.554,.48,0);
\draw[thin] (0,0,0) -- (0,.48,0) node[anchor=west]{$M$};
\node[rotate=60] at (.34,0,.73) {core axis};
\draw[<->] (.25,0,.433) arc (30:70:0.7) ;
\node[] at (0.8,0,1.5)  {$\delta$};
\node[] at (0.92,0,1.5)  {$\delta$};
\node[] at (-.14,.75,.30)  {$\theta$};
\node[] at (-.035,.0,.13)  {$\beta$};
\node[] at (-.51,.34,0)  {$Q$};
\node[] at (-.25,.34,.12)  {$r$};
\node[] at (.73,-.8,.0)  {$r$};
\node[] at (.5,.0,.7)  {$L$};
\node[] at (.07,.91,0)  {$K$};
\node[] at (1.01,0,1.8)  {$A$};
\tdplotsetrotatedcoords{0}{30}{0}
\draw[tdplot_rotated_coords] (0,-.07,0) circle [x radius=.84,y radius=.84];
\draw[tdplot_rotated_coords] (0,.48,0) -- (-.64,.48,0) node[anchor=south]{$N$};
\draw[tdplot_rotated_coords] (0,0,0) -- (-.64,.48,0);
\end{tikzpicture}
\caption{Front of inclined air shower falling on a horizontal surface.}
\label{contour2}
\end{center}
\end{figure}
\\
As for the cylinder model, the tilted circle is perpendicular to the shower direction. We take a point $N$ on the tilted circle as shown in Fig. \ref{contour2}. Its distance with respect to the origin is denoted as $r$. The projection of $N$, along the cone, on the horizontal plane is point $P$. $M$ is a point on the $y$-axis with identical $y$-coordinate as $N$. The angle between $NM$ and $QM$ is equal to the zenith angle $\theta$. The angle between $NP$ and the vertical axis is equal to $\theta +\delta$. From the geometry it follows
\begin{equation}\label{20}
(x_A, y_A, z_A)= \left( L \sin \theta, 0, L \cos \theta \right)
\end{equation}
and
\begin{equation}\label{21}
(x_N, y_N, z_N)= L \tan \delta \left( - \sin \beta \cos \theta, \cos \beta,  \sin \beta \sin \theta \right) \ ,
\end{equation}  
where
\begin{equation}\label{22}
\sin \beta = \frac{MN }{ON} 
\end{equation} 
and $ON=r$. The line through $A$ and $N$ intersects the $z=0$ plane in point $P$ with ($x$,$y$) coordinates
\begin{equation}\label{23}
(x, y)= \frac{L \tan \delta}{1-\tan \delta \sin \beta \tan \theta} \left( - \frac{\sin \beta}{\cos \theta}, \cos \beta \right) \ .
\end{equation}
As before we let $k$ be the $y$-coordinate where the projected contour intersects de positive $y$-axis. Using the coordinates as given before we find, to first order in $\delta$, the following approximate lengths of paths $AP$ and $AK$:
\begin{equation}\label{24}
AP\approx L(1+\delta \tan \theta \sin \beta) 
\end{equation}
and
\begin{equation}\label{25}
AK\approx L \ .
\end{equation}
Taking 1030 g cm$^{-2}$ for the atmospheric depth at ground level the slant atmospheric depth experienced in these paths is 
\begin{equation}\label{26}
X_{AP} \approx \frac{1030 \cdot (1+\delta \tan \theta \sin \beta )}{\cos \theta} 
\end{equation}
and
\begin{equation}\label{27}
X_{AK} \approx \frac{1030}{\cos \theta}  \ .
\end{equation} 
The difference $\Delta X$ between the atmospheric depth experienced by path $AP$ and path $AK$ is
\begin{equation}\label{28}
\Delta X =X_{AP}-X_{AK} \approx \frac{1030 \cdot \delta \tan \theta \sin \beta}{\cos \theta}  \ .
\end{equation}
This corresponds to an additional attenuation given by
\begin{equation}\label{29}
e^{- \Delta X / \lambda }  \ ,
\end{equation}
which can be elaborated to
\begin{equation}\label{30}
e^{- \Delta X / \lambda }=e^{- 7.9 \cdot 10^3  \xi  \delta  \sin \beta \tan \theta \cos^{-1} \theta} \ .
\end{equation}
For $\delta$ we have
\begin{equation}\label{31}
\delta \approx \frac{r}{L} = \frac{r \cos \theta}{h}  \ ,
\end{equation} 
where $h$ is the altitude of the apex in m. Hence
\begin{equation}\label{32}
e^{- \Delta X / \lambda }=e^{- 7.9 \cdot 10^3 h^{-1}  \xi  r  \sin \beta \tan \theta } \ .
\end{equation}
Since $r \sin \beta =MN \approx -x \cos \theta$ the attenuation can be written as 
\begin{equation}\label{33}
e^{- \Delta X / \lambda }=e^{ 7.9 \cdot 10^3 h^{-1}  \xi  x  \sin \theta} \ .
\end{equation}
Equating $\nu(x,y)$ with $\nu(0,k)$ we obtain
\begin{equation}\label{34}
e^{-(r/r_0)^w} e^{ 7.9 \cdot 10^3 h^{-1}  \xi  x  \sin \theta} =  e^{-(k/r_0)^w} \ .
\end{equation} 
The exponent of the attenuation differs only by a factor $7.9 \cdot 10^3 h^{-1}$ from the one in the previous model. Proceeding in a similar way as in the previous section will therefore lead to a shift which is $7.9 \cdot 10^3 h^{-1}$ times as large as the one of the previous model:
\begin{equation}\label{35}
x_M = \frac{7.9 \cdot 10^3}{h} \cdot \frac{\xi}{vw} \cdot k^{2-w} \cdot \frac{\tan \theta}{\cos \theta} \ .
\end{equation}
For a shower with zenith angle $60^\circ$ it is found for the difference in atmospheric depth between the late and early part of the shower $2\Delta X =370$ g cm$^{-2}$ at a distance of 1000 m from the core \cite{valino2010}. From equations (\ref{28}) and (\ref{31}) it follows for the difference between the late ($\beta = \pi /2$) and early ($\beta = - \pi /2$) part:
\begin{equation}\label{36}
2\Delta X \approx \frac{2 \cdot 1030 \cdot r \tan \theta }{h}  \ .
\end{equation}
For $2 \Delta X =370$ g cm$^{-2}$, $r=1000$ m and $\theta=60^\circ$ the latter equation is satisfied if $h=9.6 \cdot 10^3$ m. If we substitute $h=9.6 \cdot 10^3$ m, $r_0=0.025$ m$^{-1}$, $w=0.25$ and $\xi = 7.0 \cdot 10^{-4}$ m$^{-1}$, the `cone' model prediction for the shift in m is
 \begin{equation}\label{37}
x_M = 9.2 \cdot 10^{-4} \cdot k^{1.75} \cdot \frac{\tan \theta}{\cos \theta} \ .
\end{equation}
The latter is only 16\% smaller in comparison to the cylinder model prediction. It can be imagined that the energetic particles near the core point on average to an apex with larger. That would correspond to an even smaller prediction. We return to this in Section 7 where we discuss the situation for $h$ being a function of $k$. Anyway, in the following we will compare the shift solely with the cylinder model prediction ({\ref{19}), which is equal to the cone model prediction (\ref{35}) if $h \approx 7.9 \cdot 10^3$ m. 

\section{Method}
In this section we describe the method of investigation of the shift in the lateral density. The method can be applied for the electron density, muon density and the combined density of electrons and muons together. We restricted ourselves to lateral densities of proton initiated showers. The showers were generated with CORSIKA-v7.4 \cite{corsika1}, with QGSJET-II-04 \cite{ostap1,ostap2} + GHEISHA \cite{gheisha1} for the hadronic interactions. The showers were generated \emph{without thinning}. The horizontal observation level was set to 10 m. The energy cuts are 0.3 GeV for hadrons and muons and 3 MeV for electrons and photons. For each shower the lateral distribution is binned with bin size equal to $10 / \sqrt{\rho}$ m. As an illustration the binned lateral combined density of an arbitrary $10^{17}$ eV shower with $45^\circ$ zenith angle and $0^\circ$ azimuth angle is shown in Fig. \ref{density17ev45}.
\\
\begin{figure}[htbp]
\begin{center}
\includegraphics[width=10cm]{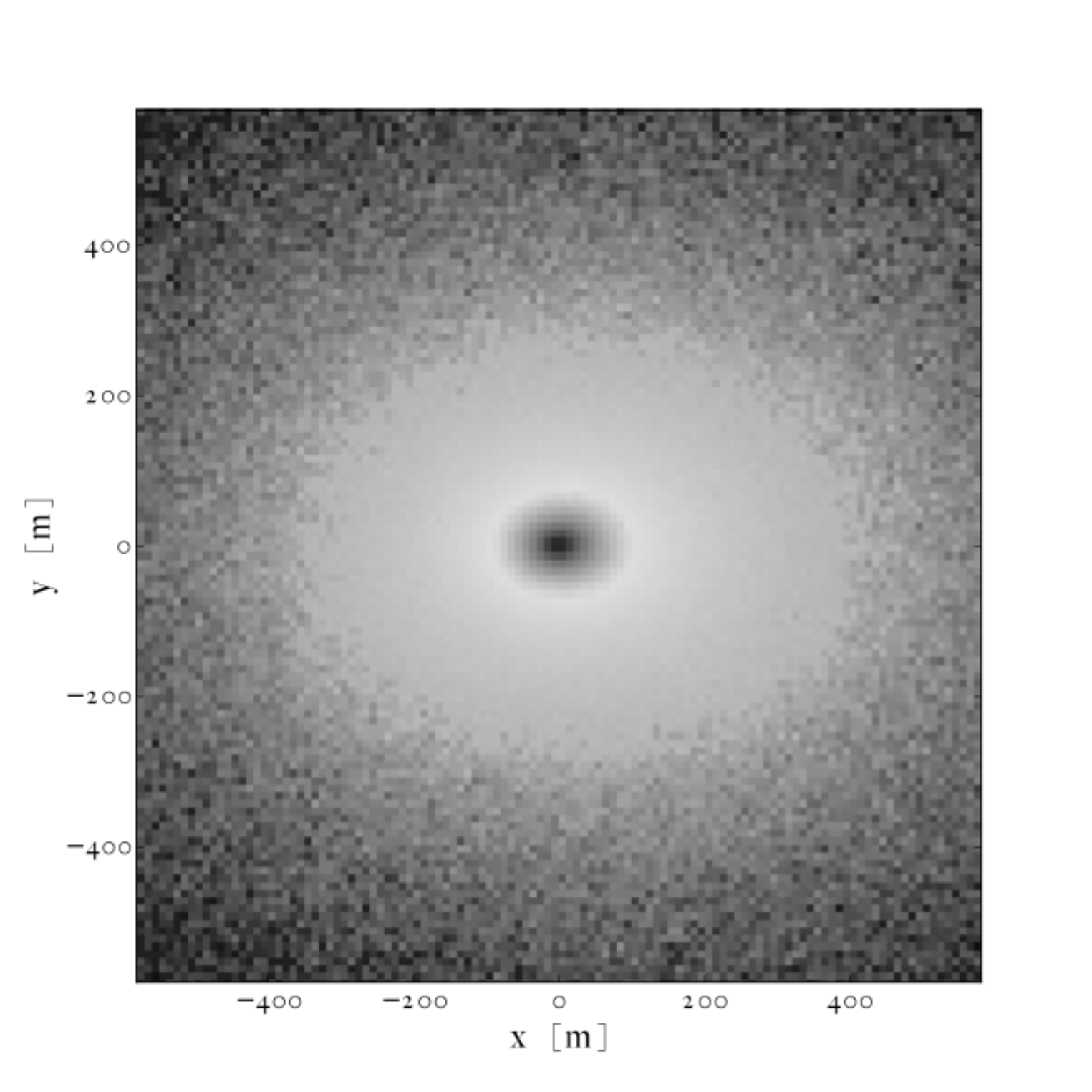}
\caption{The asymmetric lateral density of a $10^{17}$ eV proton shower with zenith angle $45^\circ$ and azimuth angle $0^\circ$.}
\label{density17ev45}
\end{center}
\end{figure}
\\
From the binned density, smoothened by means of a Gaussian filter with one bin ($x \times x$) as sigma, the iso-density contours are determined. By means of minimization of the sum of squares the contours are fitted by an ellipse with equation
\begin{equation}\label{38}
\left(\frac{x -x_M}{a} \right)^2 + \left(\frac{y -y_M}{b} \right)^2  =1 \ ,
\end{equation}
where $a$ and $b$ are the semi-major and semi-minor axes respectively. For the example shower of Fig. \ref{density17ev45} the final contour with density $< \rho > = 1$ m$^{-2}$ is shown in Fig. \ref{contourellips} together with the ellipse resulting from the fit. In this way we obtain values for the semi-major axis, the semi-minor axis and the value of $x_M$. The center of the ellipse is denoted as $M$. The shower core is at the origin $O$. The focal points $F_1$ and $F_2$ are at distance $c$ from the center $M$. This distance is related to the semi-major and semi-minor axis via $c^2=a^2-b^2$. 
\\
\begin{figure}[htbp]
\begin{center}
\includegraphics[width=10cm]{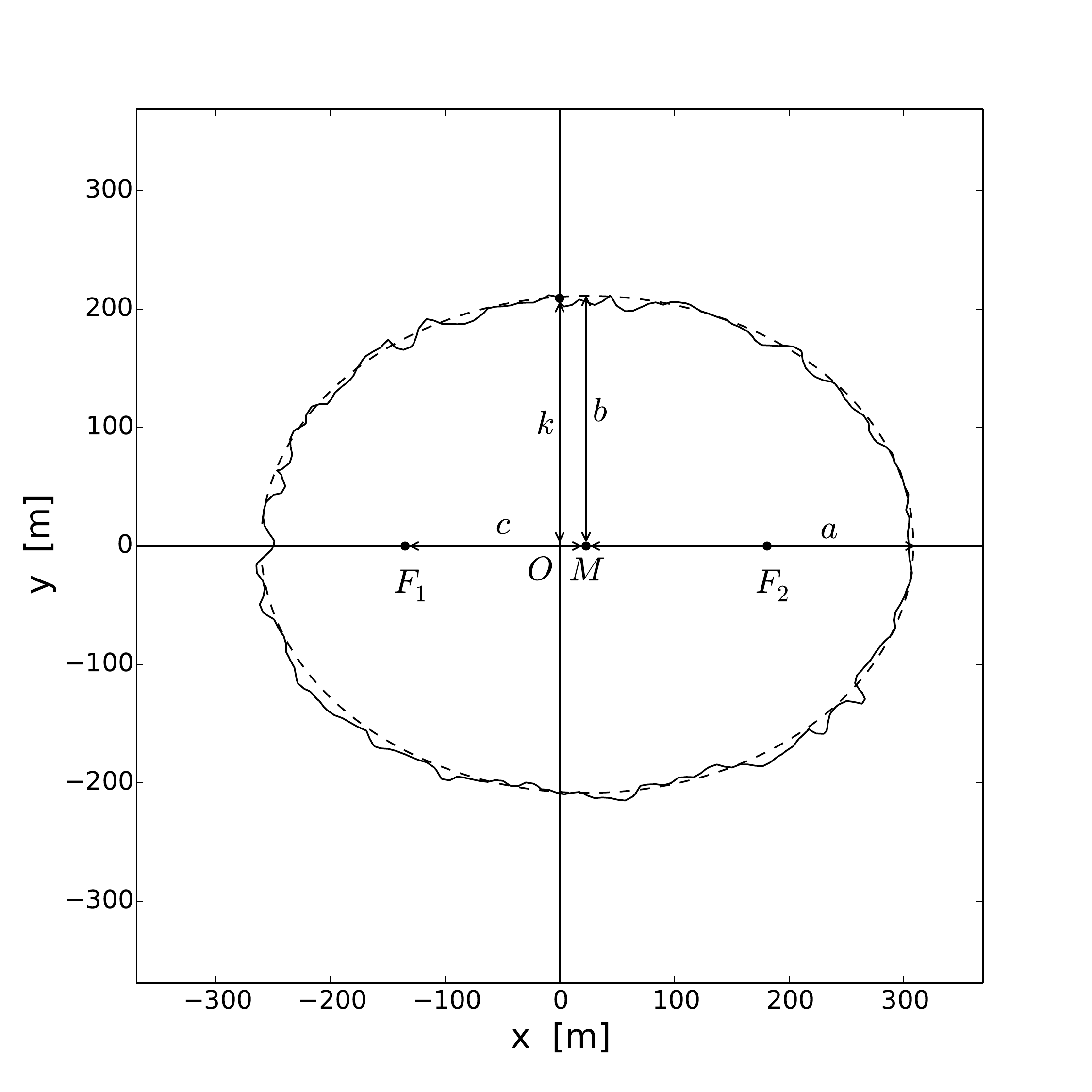}
\caption{The iso-density contour for density $\rho = 1$ m$^{-2}$ of a $10^{17}$ eV proton shower with zenith angle $45^\circ$ and $0^\circ$ azimuth angle (solid). It is with good approximation equal to an ellipse (dashed).}
\label{contourellips}
\end{center}
\end{figure}
\\
Next to $x_M$ the fit also delivers a value for the $y_M$ coordinate. Its value, which is close to zero as it should, will be left out of the analysis. Ignoring $y_M$ the Eq. (\ref{38}) can be written either as 
\begin{equation}\label{39}
\left( (x -x_M)\frac{b}{a} \right)^2 + y^2  =b^2 
\end{equation}
or as 
\begin{equation}\label{40}
\left( x^2 - 2 x \cdot x_M \right)    \frac{b^2}{a^2}  + y^2  =k^2 \ ,
\end{equation}
where
\begin{equation}\label{41}
k^2 = b^2-\frac{b^2}{a^2} x_M^2 \ .
\end{equation}
So, having determined the semi-major axis $a$, the semi-minor axis $b$ and the shift $x_M$, we also know the corresponding value of $k$.
\\ \\
From the model analyses, we expect $b=a \cos \theta$. In Fig. \ref{ellipticity} the value of $b/a$ is plotted against zenith angle for the combined density of the simulated showers. 
\begin{figure}[htbp]
\begin{center}
\includegraphics[width=8.5cm]{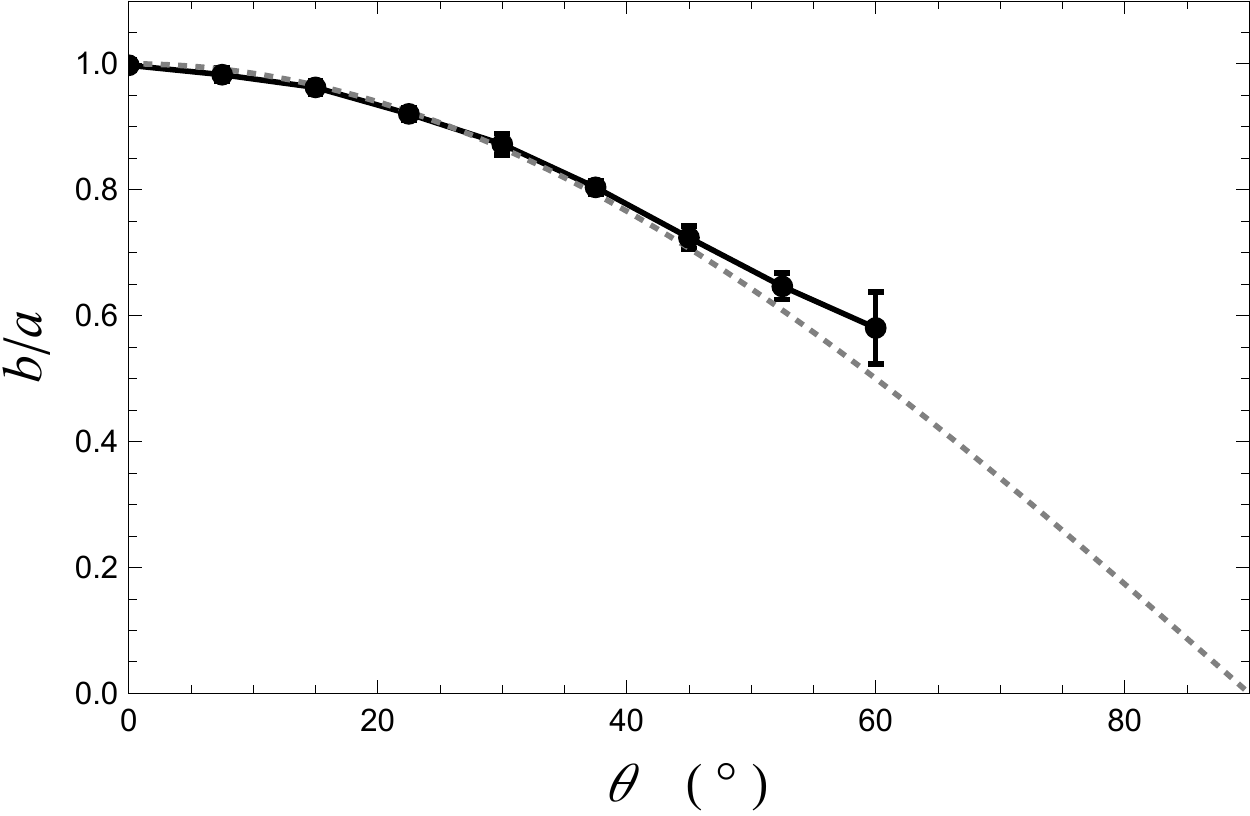}
\caption{The ratio of the semi-minor and semi-major axis of elliptic lateral combined density of simulated showers for different zenith angles. The dashed line is $\cos \theta$.}
\label{ellipticity}
\end{center}
\end{figure}
\\
We see $b/a$ follows $\cos \theta$ for small zenith angles. For larger zenith angles a small deviation shows up. The deviation increases with zenith angle. Since the expectation $b=a \cos \theta$ is based on polar symmetric iso-density contours in the front plane, the deviation suggests the iso-density contours in the front plane to be slightly elliptic with the major axis perpendicular to the azimuth direction. With the substitution of $\cos \theta$ for $b/a$ the Eqs. (\ref{39}) through (\ref{41}) reduce to the Eqs. (\ref{5a}) through (\ref{5c}).
\\ \\
The method has been tested with artificial Poisson randomized shifted elliptic densities in order to check if the imposed shift is returned. The deviations between the imposed and returned shifts were small, around 1 m for density of 1 m$^{-2}$ or less. Next to the inaccuracy of the method there also are contributions to the deviations due to the fluctuations of the densities. As a measure for the uncertainty the deviations of $y_M$ with respect to the expected value 0 are taken. For the simulated showers we found the standard deviation of $y_M$ to depend on density roughly as $\sigma_y \approx \rho^{-1/3}$. Assuming the standard deviation of $x_M$ to be comparable, we use it for the size of the error bars in the diagrams in the next section. 

\section{General Monte Carlo results}
In this section we consider the lateral densities of two proton initiated showers with energy 100 PeV and 10 PeV both with zenith angle $45^\circ$. For both showers the shifts were determined for the electron density, the muon density and the combined density. For the combined density and the muon density the shift was determined for densities 0.001, 0.002, 0.003, 0.004, 0.006, 0.008, 0.01, 0.02, 0.03, 0.04, 0.06, 0.08, 0.10, 0.20, 0.30, 0.40, 0.50, 0.64, 0.81, 1.00, 1.44, 2.00 and 5.0 m$^{-2}$. For the electron density the same densities were used with the densities 0.0002, 0.0003, 0.0004, 0.0006 and 0.0008 m$^{-2}$ added to that. The shifts are shown in Fig. \ref{explainshift45} respectively Fig. \ref{explainshift2}. The dashed curve in Figs. \ref{explainshift45} and \ref{explainshift2} is the model prediction (\ref{19}). 
\begin{figure}[htbp]
\begin{center}
\includegraphics[width=8cm]{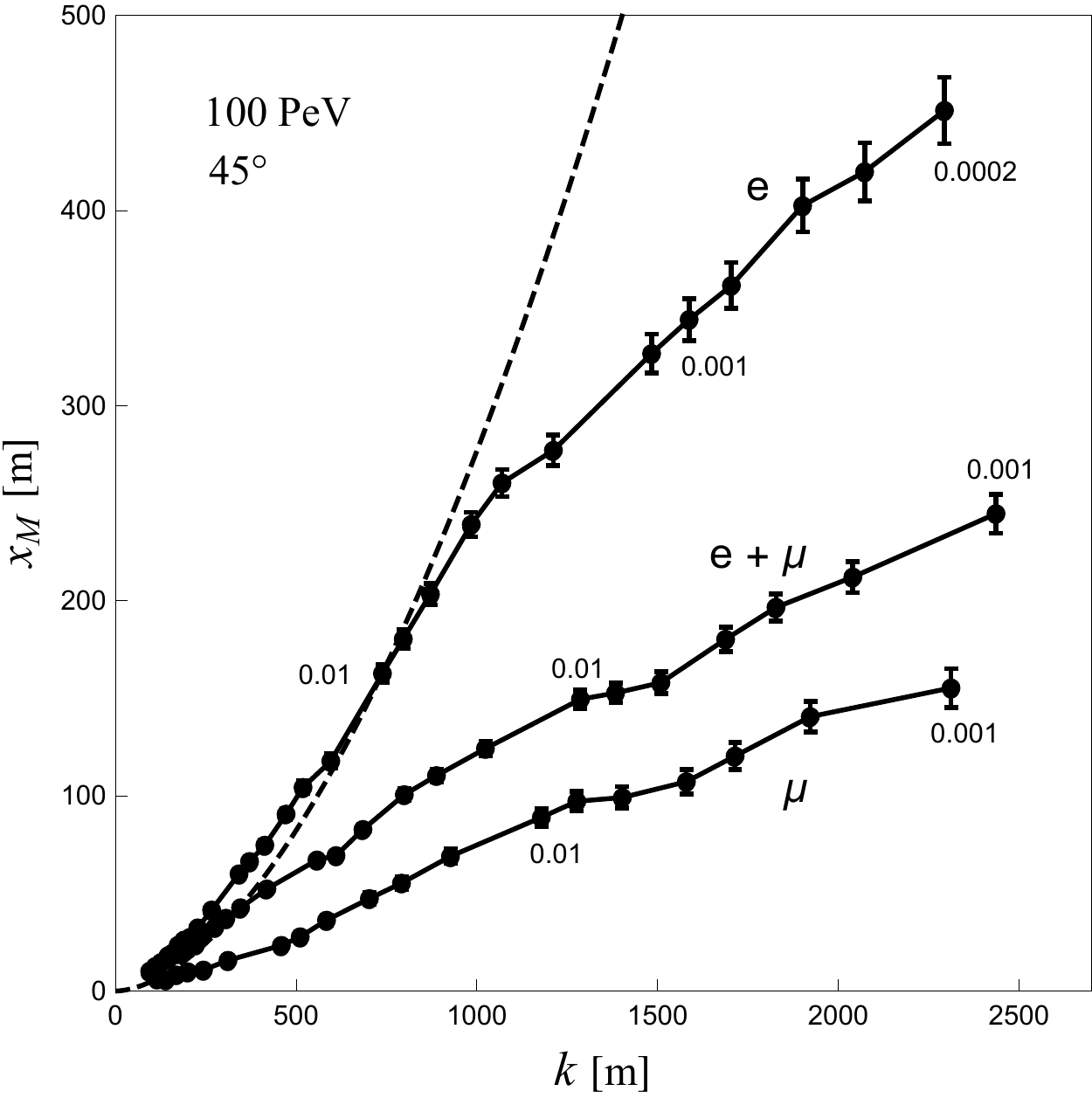}
\caption{The shifts of the electron, muon and combined density of a 100 PeV shower with zenith angle $45^\circ$. The data points are connected with line pieces to guide the eye. The dashed curve is the model prediction. The densities 0.01, 0.001 and 0.0002 are depicted.}
\label{explainshift45}
\end{center}
\end{figure}
\\
Comparing the model prediction with the determined shifts of the electron density we see the model prediction follows to a certain extent the shifts as determined for the electron density. The model predicts too low for $k<700$ and $k<500$ m for the 100 PeV reps. 10 PeV shower.
\\ \\
There are many reasons for the model to deviate from the determined shifts. To begin with, the plotted model prediction for the shift was based on a constant value for $h$. In reality $h$ will depend on the distance to the shower core, as visualized in Fig. 2 of \cite{garcia-pinto2009}. This suggests a large value for $h$ near the core and a decreasing value for $h$ for increasing $k$. According to the cone model a larger value for $h$ implies a smaller value for the shift. As a consequence it will enhance the underestimation near the core. A small effect has the atmospheric depth decreasing exponentially with altitude. This will flatten the model curve for large $k$. We just took a constant value for the atmospheric depth in the model which is sufficient for our region of interest: $k<1000$ m and $\theta < 60^\circ$. 
\begin{figure}[htbp]
\begin{center}
\includegraphics[width=8cm]{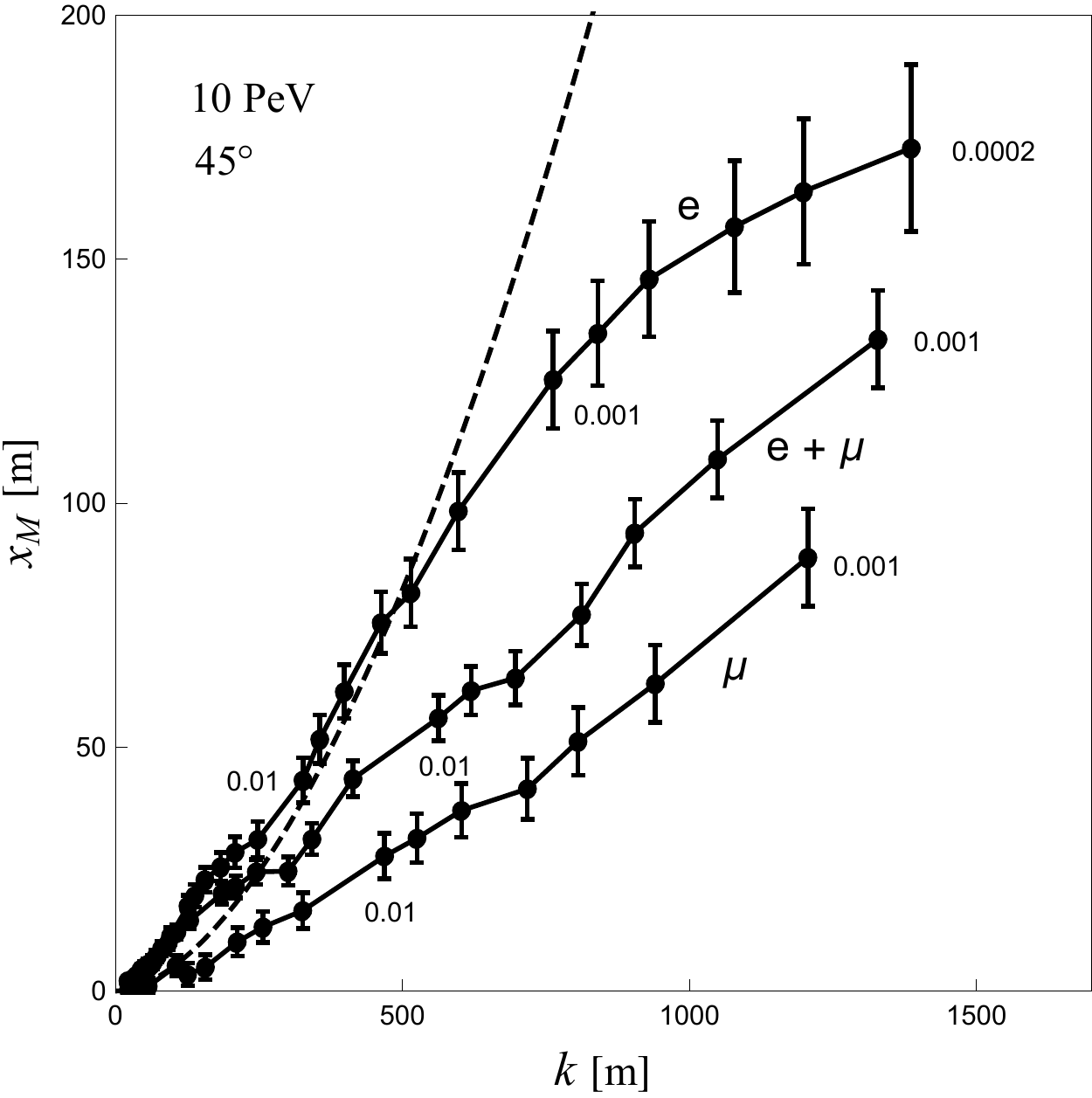}
\caption{The shifts of the electron (triangle), muon (diamond) and combined (circle) density of a 10 PeV shower with zenith angle $45^\circ$. The dashed curve is the model prediction. The numbers 0.01, 0.001 and 0.0002 depict the density.}
\label{explainshift2}
\end{center}
\end{figure}
\\ 
The most important reason for the bad prediction probably is the local variation of the attenuation. It can be imagined that the attenuation is large near the core and decreases for increasing $k$. The latter would enhance the shift near the core and flatten the model curve further away from the core. Alternatively, it might bring the model curve more in agreement with the shift curve of the combined density. To model it requires the knowledge of $\xi(k)$ as a function of $k$. The local $h(k)$ can possibly be obtained from simulated showers by inspection of the directions of the electrons when they arrive at the observation plane. The knowledge of $\xi(k)$ seems more difficult: besides the inspection of the local energy distribution of electrons it also requires a relation between the distributions and the local attenuation. On the other hand, if one succeeds in describing and modeling the atmospheric depth and $h$ as functions of $k$, parameterized by zenith angle, there is an opportunity to retrieve the local attenuation $\xi(k)$ from the shift $x_M(k)$ as determined from the simulated electron density. For a model which predicts the shifts of the muon density one has to consider the decay of muons to electrons and the subsequent atmospheric attenuation of the electrons. The combined density the shift then follows from
\begin{equation}\label{41a}
x_{M,e+\mu}(k) =\frac{\rho_e x_{M,e}(k)+\rho_\mu x_{M,\mu}(k)}{\rho_e (k)+\rho_\mu (k)} \ .
\end{equation}
Because of the aforementioned reasons it is difficult to derive a precise model for the shifts of the electron density, let alone for the muon density and the combined density. Therefore we will not proceed in that direction. Instead, we will focus our attention on the behavior of the determined shift for relatively large densities. 
\\ \\
In Fig. \ref{shiftbranche} the shift curves of the combined density of a 10 PeV and a 100 PeV shower with zenith angle $45^\circ$ are once more plotted. They are identical to the ones in Fig. \ref{explainshift45} and Fig. \ref{explainshift2}, except that the dots and error bars are left. 
\begin{figure}[htbp]
\begin{center}
\includegraphics[width=9cm]{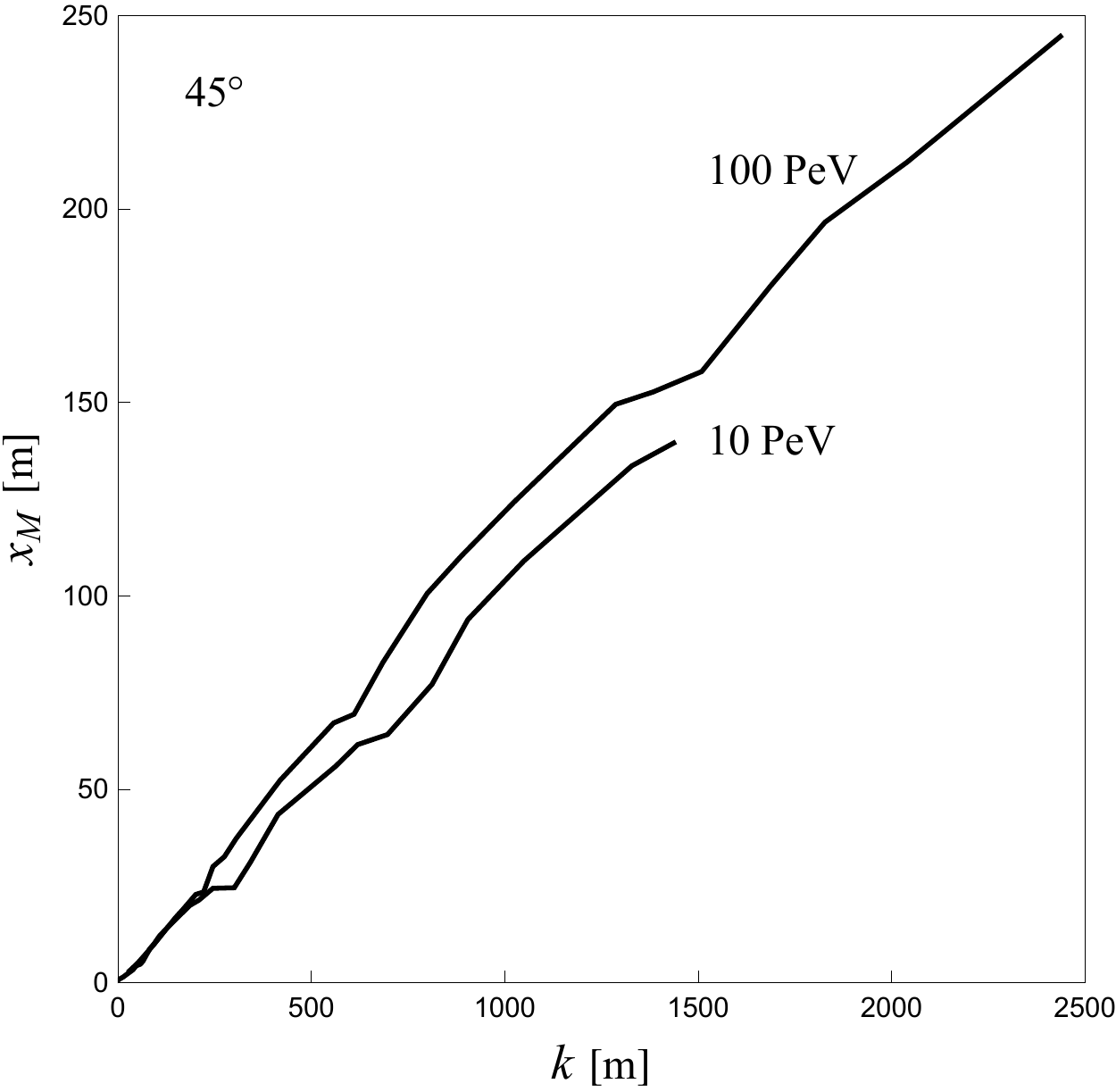}
\caption{The shifts of the combined density of a 10 PeV shower and a 100 PeV shower both with zenith angle $45^\circ$.}
\label{shiftbranche}
\end{center}
\end{figure}
\\
The slope of both curves show some curious irregularities. For the 100 PeV curve these are around $k=200$, 580 and 1400 m. For the 10 PeV curve we see them around $k=250$ and 650 m. The question arises whether these irregularities are the remnants of consecutive hadronic interactions.
\\ \\
The two shift curves fall on top of each other for $k<200$ m. At $k=200$ m they do branch. Beyond the fork the difference between the curves slightly increase for increasing $k$. In general this means that the shift curves are not independent of shower size. The density of the 10 PeV shower at the branching point is about 0.06 m$^{-2}$. In Fig. \ref{explainshift3} the shifts of the combined density as found for the 10 PeV shower is plotted on top of the ones for the 100 PeV shower. For the 10 PeV shower the plotted densities (white) are 0.2, 0.3, 0.4, 0.5, 0.64, 0.81, 1.00, 1.44, 2.0, 5.0 and 10 m$^{-2}$. For the 100 PeV shower the plotted densities (black) are 0.3, 0.4, 0.5, 0.64, 0.81, 1.00, 1.44, 2.0, 5.0, 10, 20 and 50 m$^{-2}$.
\begin{figure}[htbp]
\begin{center}
\includegraphics[width=10cm]{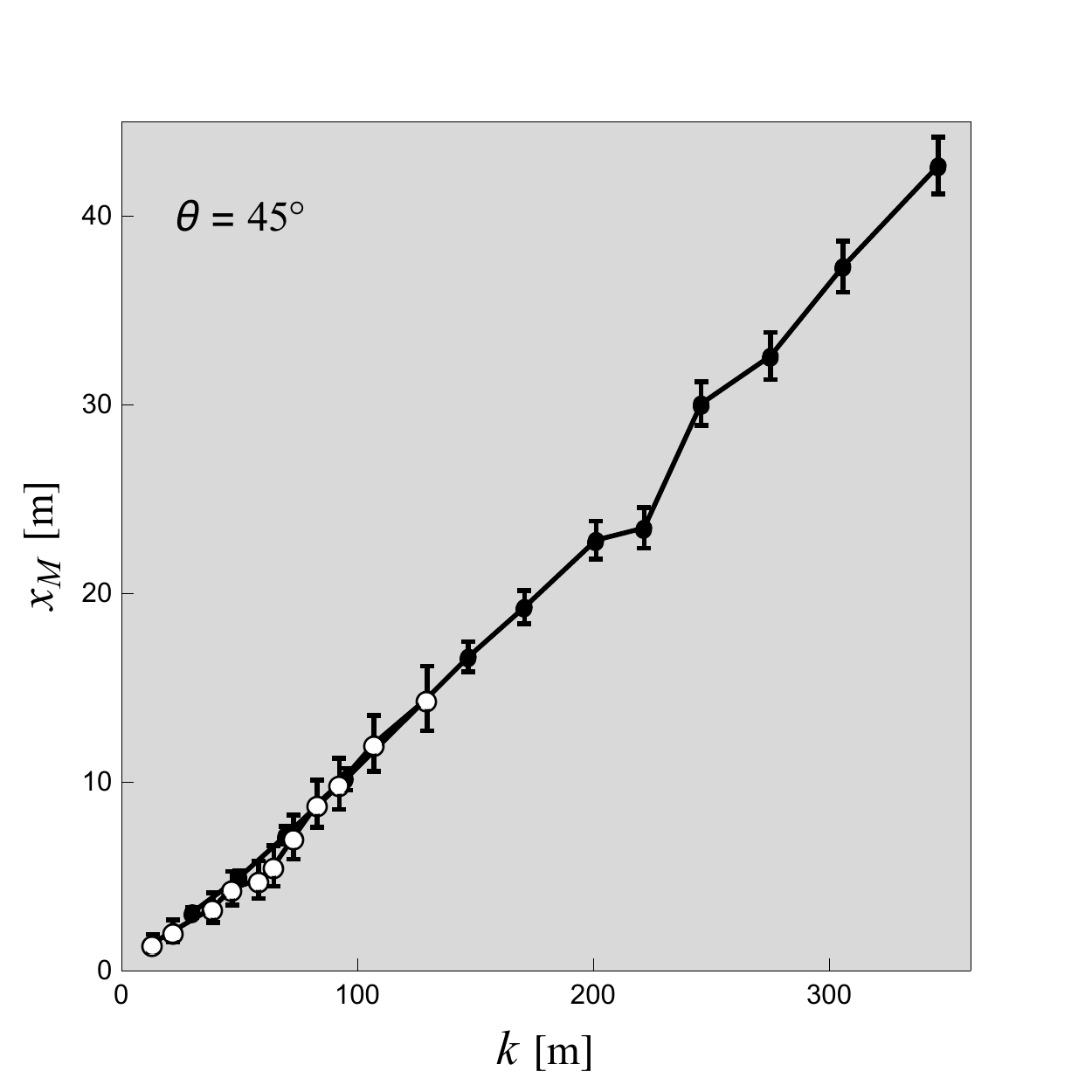}
\caption{The shifts of the combined density for a 10 PeV shower (white) on top of the ones for a 100 PeV shower (black), both showers with zenith angle $45^\circ$.}
\label{explainshift3}
\end{center}
\end{figure}
\\
We see the shift curves practically fall on top of each other within the given density domains. This means that we can try to find a relation between $x_M$ and $k$ independent of shower size (or energy) similar to the model prediction. In addition, the curves are almost linear. As we will see further on the latter allows for an analytical solution for the shifted polar density. 

\section{Specific Monte Carlo results}
In this section we will investigate the shift in combined densities of a set of simulated showers. The energies of the showers are $10^{15}$, $10^{16}$, $10^{17}$ and $10^{18}$ eV. The zenith angles of the showers range from $7.5^\circ$ through $60^\circ$, in steps of $7.5^\circ$. As an illustration the ratio $N_e / ( N_e + N _\mu )$ is plotted against energy for several zenith angles in Fig. \ref{eldivelmu}. The energy - zenith angle entries are shown as black dots. 
\begin{figure}[htbp]
\begin{center}
\includegraphics[width=10cm]{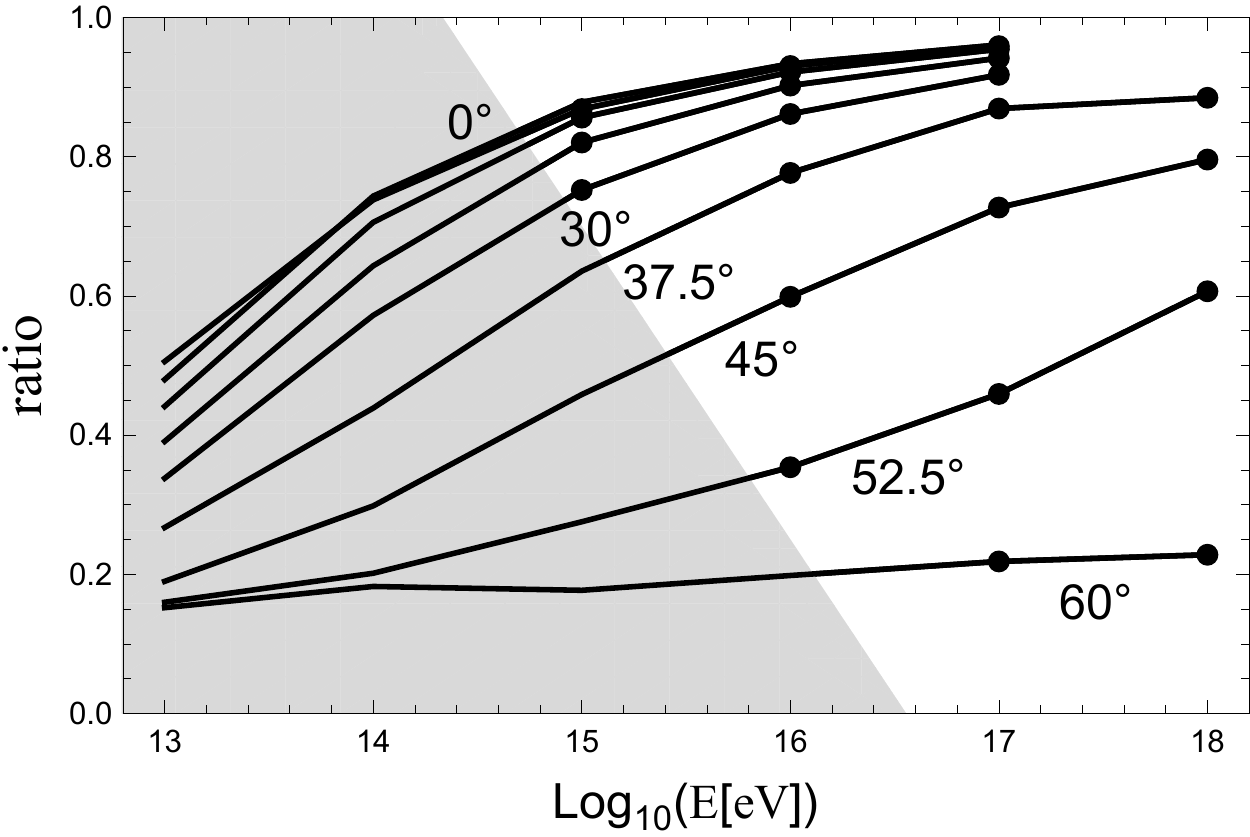}
\caption{The ratio $N_e / (N_e + N_\mu )$ against energy for zenith angles $0^\circ$ through $60^\circ$, in steps of $7.5^\circ$.}
\label{eldivelmu}
\end{center}
\end{figure}
\\
Not all possible combinations of energies and zenith angles are applicable for investigation. For  showers with relatively low energy and relatively large zenith angle, shown as the gray region in Fig. \ref{eldivelmu}, the small shower size at observation level does in general not allow for a determination of the shift. The choice for generating showers without thinning is made to avoid possible deviations caused by thinning. The consumption of computer time and of storage space grows exponentially with the size of the simulated shower  \cite{apcworkshop}. This is extremely the case for shower simulation without thinning. As a consequence the library of showers generated without thinning is limited, in particular for large energies. For the largest energy considered, $10^{18}$ eV, the library is momentarily limited to 10 showers for zenith angle $60^\circ$, 10 for zenith angle $52.5^\circ$, 8 for zenith angle $45^\circ$, 5 for zenith angle  $37.5^\circ$ and none for zenith angles $30^\circ$, $22.5^\circ$ and $15^\circ$. To obtain a sort of equal share in our diagrams we take 10 showers for each of the other energy - zenith angle entries. For each shower we determine the iso-density contours for combined densities 0.50, 0.64, 0.81, 1.00, 1.44, 2.0, 5.0, 10, 20 and 50 m$^{-2}$ for as far as these densities occur in a shower, thus maximum 10 data points per shower. For the densities considered this is close to the shift as we would have obtained it from the electron density, except for regions were the muon component dominates: for large distances to the core and for zenith angles in the neighborhood of $60^\circ$ and larger. As depicted in Fig. \ref{eldivelmu} for zenith angle 7.5$^\circ$ through 30$^\circ$ showers were used with energy $10^{15}$, $10^{16}$ and $10^{17}$, for zenith angle 37.5$^\circ$ through 52.5$^\circ$ showers were used with energy $10^{16}$, $10^{17}$ and $10^{18}$ eV and for zenith angle 60$^\circ$ showers were used with energy $10^{17}$ and $10^{18}$ eV. For 10 showers at 3 energy decades we obtained a maximum of 300 data points for each zenith angle. 
\\ \\
As we will see, and as suggested by the model results, for each zenith angle the data points follow a curve independent of energy. The curves are fitted with a function similar to the one resulting from the models. To be specific, for each zenith angle we plot the $x_M$ against $k$ and fit the result by the equation
\begin{equation}\label{43}
x_M  = Ak^B \cdot \frac{\tan \theta}{\cos \theta}  \ .
\end{equation}
As an illustration the $x_M$ are plotted against $k$ for zenith angle $30^\circ$ and fitted with Eq. (\ref{43}), see Fig. \ref{xmvsk30}. 
\begin{figure}[htbp]
\begin{center}
\includegraphics[width=8cm]{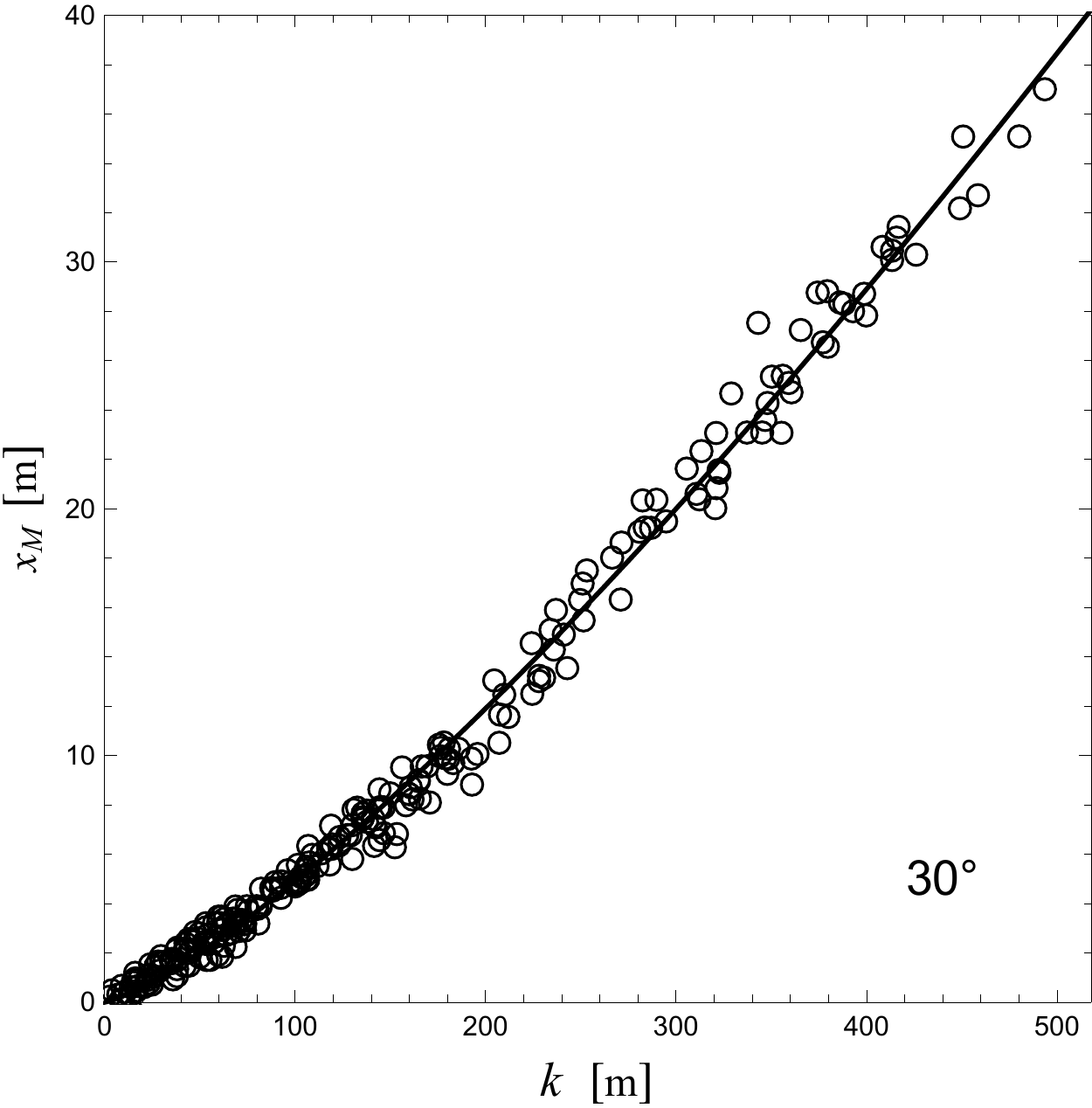}
\caption{The shift $x_M$ against $k$ (unfilled circles) and the best fit curve for zenith angle $30^\circ$.}
\label{xmvsk30}
\end{center}
\end{figure}
\\
In the latter figure the error bars since they are in most of the cases smaller than the size of the plot markers. The $x_M$ grows with $k$ along a curve independent of the primary energy of the showers. For the 1 PeV showers the data points are in the region $k<80$ m. For the 10 PeV and 100 PeV the regions are $k<250$ m and $k<500$ m respectively. For the parameters of the fit we find $A=0.020$ and $B=1.28$. For the goodness of fit we find for the Pearson $\chi^2$ test 0.98 as the $p$-value. These figures hold for zenith angle $30^\circ$. For zenith angle 7.5$^\circ$ through 30$^\circ$ the diagrams are shown in Fig. \ref{xmvsklinalla}. For zenith angle 37.5$^\circ$ through 60$^\circ$ the diagrams are shown in Fig. \ref{xmvsklinallb}. 
\begin{figure}[htbp]
\begin{center}
\includegraphics[width=10.5cm]{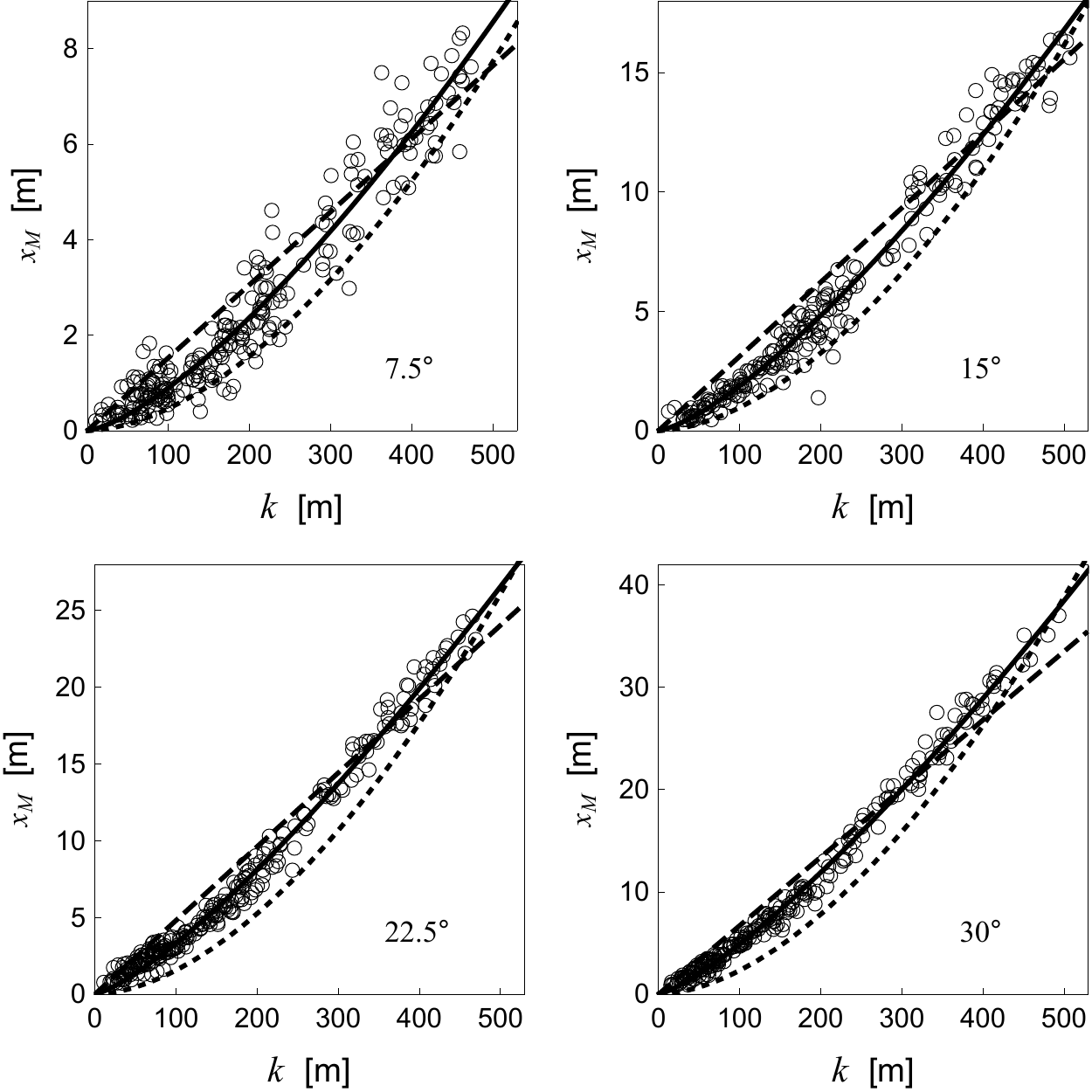}
\caption{The shift $x_M$ against $k$ (unfilled circles), the best fit curve (solid), the model prediction (dotted) and a linear curve (dashed) for  zenith angle 7.5$^\circ$ through 30$^\circ$.}
\label{xmvsklinalla}
\end{center}
\end{figure}
\begin{figure}[htbp]
\begin{center}
\includegraphics[width=10.5cm]{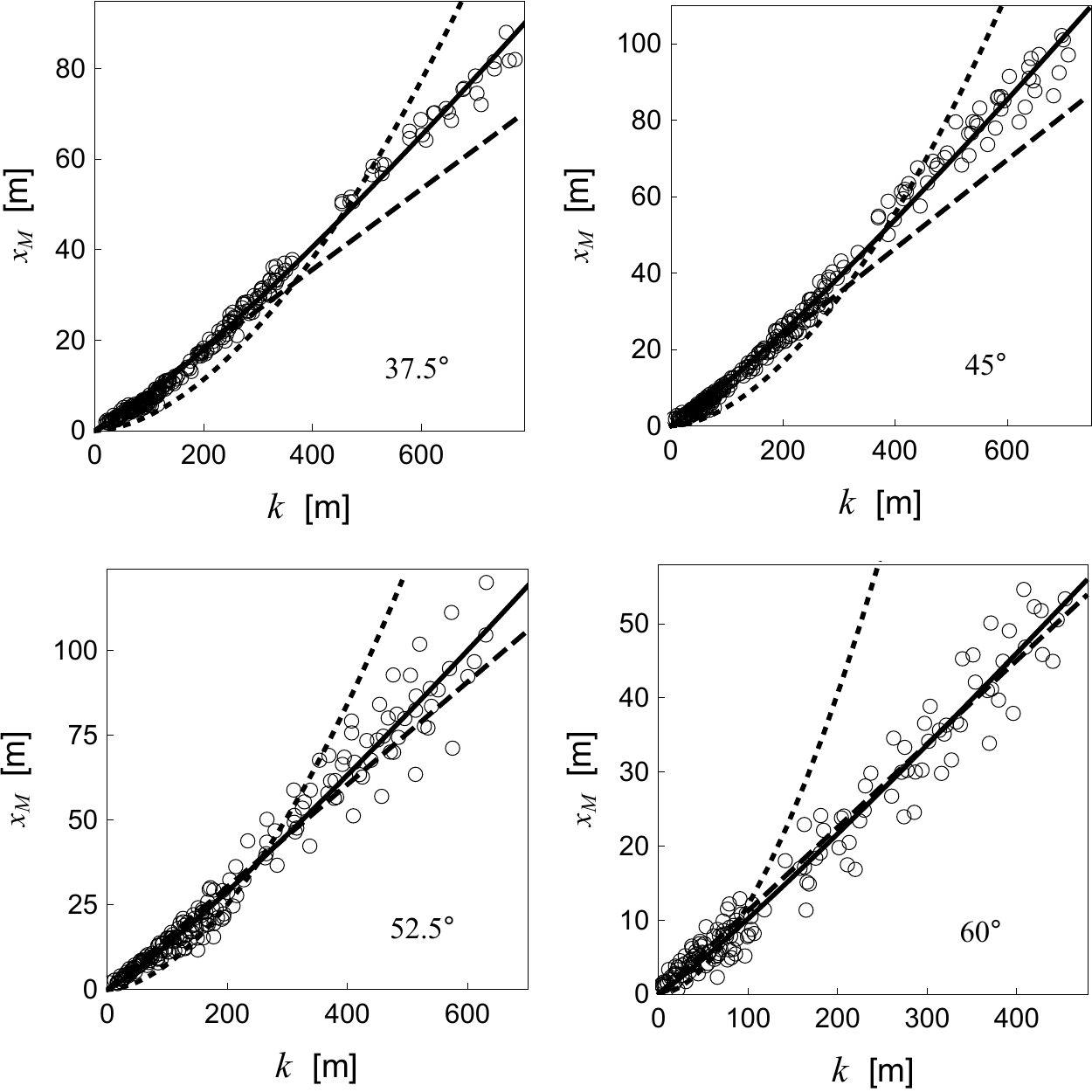}
\caption{The shift $x_M$ against $k$ (unfilled circles), the best fit curve (solid) and the model prediction (dotted) and a linear curve (dashed) for zenith angle 37.5$^\circ$ through 60$^\circ$.}
\label{xmvsklinallb}
\end{center}
\end{figure}
\begin{table}[htbp]
\renewcommand{\arraystretch}{1.4}
\setlength{\tabcolsep}{6pt}
\begin{center}
\begin{tabular}{ | c | c | c | c | c | c |} \hline
$\theta$ & $A$ & $B$ & $\chi^2$ & $p$-value & datasize \\ \hline
7.5$^\circ$  & 0.010 & 1.41 & 25.7 & 0.11 & 248  \\ \hline
15$^\circ$  & 0.013 & 1.36 & 8.74 & 0.95 & 220  \\ \hline
22.5$^\circ$  & 0.019 & 1.29 & 11.7 & 0.86 & 274  \\ \hline
30$^\circ$  & 0.020 & 1.28 & 7.79 & 0.98 & 253  \\ \hline
37.5$^\circ$  & 0.037 & 1.17 & 12.5 & 0.82 & 250  \\ \hline
45$^\circ$  & 0.043 & 1.13 & 10.0 & 0.93 & 267  \\ \hline
52.5$^\circ$  & 0.034 & 1.13 & 9.21 & 0.91 & 199  \\ \hline
60$^\circ$  & 0.019 & 1.09 & 11.1 & 0.74 & 156  \\ \hline
\end{tabular}
\caption{Parameters $A$ and $B$ of the fits, the $\chi^2$ and $p$-values and the number of inspected contours for zenith angles as given in the first column.}
\label{abparam}
\end{center}
\end{table}
\\
In both diagrams the model predictions for the shift of the electron densities are plotted as well for reasons of comparison; for small zenith angles the combined distribution is dominated by electrons.
\\ \\
For small zenith angles, 7.5$^\circ$ and 15$^\circ$, the spread of the shifts are mainly governed by the uncertainty of the measurement. For large zenith angles, 52.5$^\circ$ and 60$^\circ$, the spread of the shifts are mainly due to shower to shower variations. For each zenith angle the values $A$, $B$, $\chi^2$ and the $p$-value have been tabulated, see Table \ref{abparam}.
\\ \\
The values of $B$ as given in Table \ref{abparam} decrease for increasing zenith angle. The values of $B$ being close to unity suggests to consider a linear relation between $x_m$ and $k$. From fits with
\begin{equation}\label{44a}
x_M  =C \cdot k \cdot \frac{\tan \theta}{\cos \theta} 
\end{equation}
it is found that $C$ scales as $\cos \theta$. On average $C\approx 0.116 \cos \theta$, except for $\theta=60^\circ$ where $C \approx 0.075 \cos \theta$. Writing $C$ as $2S \cos \theta$, the proposed linear relation is as follows
\begin{equation}\label{44}
x_M = 2S \cdot k \tan \theta  \ ,
\end{equation}
where $S=0.058$. For $\theta=60^\circ$ the value of $S$ is about 35\% smaller. The linear relation is shown in Figs. \ref{xmvsklinalla} and \ref{xmvsklinallb} as a dashed curve. For small zenith angles the linear approximation overestimates the shift for $k \approx 150$ m, the difference being just about 1 m. For zenith angle 37.5$^\circ$ and 45$^\circ$ it underestimates by 20\% in the region where the density is small, $\rho <1$ m$^{-2}$. Taking the inaccuracies for granted, a linear equation is advantageous since, as we will see further on, it allows for an analytical solution for the description of the LDF-A. At the end of Section 10 a remark will be made about the possible application of the more accurate power law (\ref{43}).

\section{The polar density}
In this section we will perform the conversion of an LDF to an LDF-A. Substitution of the shift (\ref{44}) in Eq. (\ref{5c}) gives
\begin{equation}\label{58}
\left( x^2 - 4 x \cdot S \cdot k \tan \theta \right)  \cos^2 \theta  + y^2  =k^2 \ .
\end{equation}
Solving for $k$ we obtain
\begin{equation}\label{58a}
k=-S x  \sin (2\theta) +  \sqrt{y^2 + x^2 (1+4S^2 \sin^2 \theta) \cos^2 \theta }  \ ,
\end{equation}
For $S=0.058$ the term $4S^2 \sin^2 \theta$ is negligible with respect to 1. With good approximation we therefore have
\begin{equation}\label{58b}
k=-S x  \sin (2\theta) +  \sqrt{y^2 + x^2  \cos^2 \theta }  \ ,
\end{equation}
According to Eq. (\ref{8}) the polar density $\nu (x,y)$ including the shift is obtained in Cartesian coordinates by substituting expression (\ref{58b}) for $k$ in the polar symmetric density $\rho$ and by multiplying it by $\cos \theta$.
The horizontal polar density $\nu$ can also be written in polar coordinates. With the substitution of $x=r\cos\alpha$ and $y=r\sin\alpha$ the Eqs. (\ref{58}) and (\ref{58b}) respectively read
\begin{equation}\label{59}
r^2 \cos^2 \alpha \cos^2 \theta - 2 r \cdot S \cdot k \cos \alpha  \sin (2 \theta)  + r^2 \sin^2 \alpha  =k^2 
\end{equation}
and
\begin{equation}\label{61}
k=-S r \cos \alpha \sin (2\theta) + r \sqrt{1-\cos^2 \alpha \sin^2 \theta }  \ .
\end{equation}
The second term, the square root part, is due to the ellipticity of the density as caused by the projection \cite{cillis}. The first term on the right hand side of Eq. (\ref{61}) is due to the shift. The polar density including the shift is obtained in horizontal polar coordinates in the same way as for Cartesian coordinates:
\begin{equation}\label{62}
\nu(r,\alpha)=\rho(k) \cos \theta   
\end{equation} 
with $k$ as given by (\ref{61}).
\\ \\
To obtain the polar density $\nu$ we need the polar symmetric density $\rho$. A good approximation for $\rho$ is found by polar averaging the horizontal density and fitting it with a suitable LDF. For electromagnetic showers a well known LDF is the one of Nishimura, Kamata and Greisen (NKG) \cite{K&N,Greisen}. Most LDF's are modifications of the NKG function \cite{N&W, Apel}. For muons a well known lateral density function is the one of Vernov \cite{Vernov1968}. However, it can also be described by a NKG type of function \cite{Greisen1960, Antoni2001}. For radii smaller than about 300 m the combined density of electrons and muons can also be described by an NKG type of LDF:  
\begin{equation}\label{63}
\rho_{e+\mu} (r)=N_{e+\mu} \cdot c \cdot f(r) \ ,
\end{equation}
where
\begin{equation}\label{64}
f(r)=\left(\frac{r}{r_0}\right)^{s_1}\left(1+\frac{r}{r_0}\right)^{s_2} 
\end{equation}
is the structure function and where $c$ usually is the normalization.
Formally this LDF is similar to the one used for the KASCADE experiment  \cite{Apel}. There the quantities $s-\alpha$ and $s-\beta$, with $s$ the shape parameter (a remnant of the age parameter), play a similar role as $s_1$ respectively $s_2$. The parameter $r_0$ plays a similar role as the Moli\`ere radius in the original NKG function. From the simulated showers it is found that $r_0$ is close to $30$ m. Fixing $r_0$ to 30 has only a marginally effect on the fit values for $s_1$ and $s_2$.  
\\ \\
For radii larger than about 300 m it underestimates the combined density. The deviation is caused by the relatively large muon component. To adjust for the muon component we let us motivate by the Greisen function \cite{greisen1960}. That is, we multiply the LDF  by $(1+\frac{r}{11.4 \cdot r_0})$.  The latter multiplication complicates the normalization. We therefore take $r_0=30$ and use $c$ as a fit parameter like $s_1$ and $s_2$. Thus
\begin{equation}\label{67}
\rho_{e+\mu}=N_{e+\mu} \cdot c \cdot \left( \frac{r}{30} \right)^{s_1} \left( 1+ \frac{r}{30} \right)^{s_2} \left( 1+\frac{r}{340} \right) \ .
\end{equation}

\section{Comparison with simulated densities}
The performance will be illustrated by means of the same three showers \textbf{a}, \textbf{b} and \textbf{c} as already used in Section 4. Their polar averaged combined densities and their fitting curves are plotted in Fig. \ref{ldftripelmu}. As before the polar averaged densities are binned with bin-width 1 m.
\begin{figure}[htbp]
\begin{center}
\includegraphics[width=10cm]{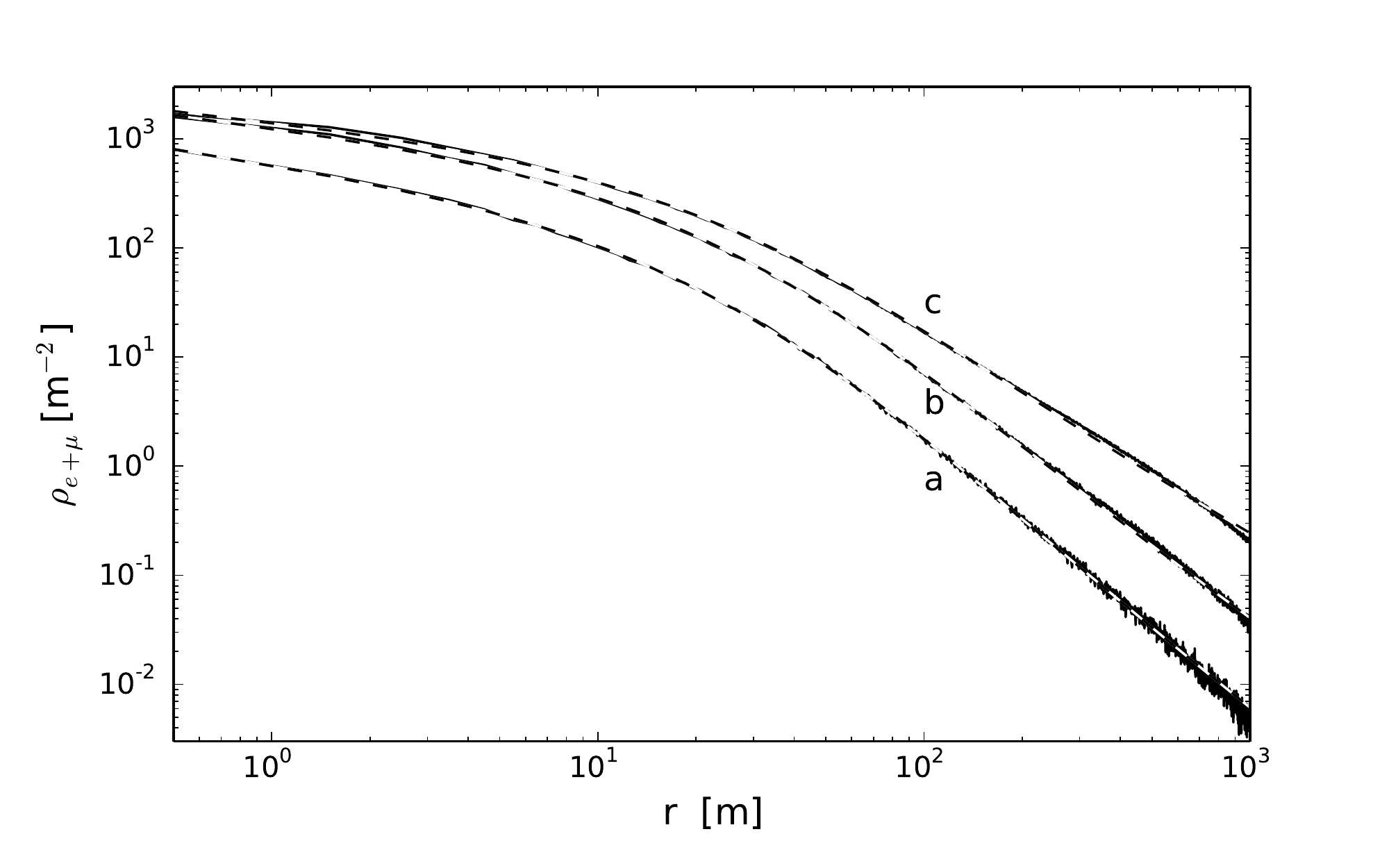}
\caption{Polar averaged combined densities for the three showers \textbf{a}, \textbf{b} and \textbf{c} as given in the text. The dashed curves are the fits with the LDF as given in the text.}
\label{ldftripelmu}
\end{center}
\end{figure}
\\ 
We see the fit curves follow the combined densities also beyond 300 m. The number of electrons and muons and the values found for the parameters are shown in Table \ref{elmuparam}.
\begin{table}[htbp]
\renewcommand{\arraystretch}{1.4}
\setlength{\tabcolsep}{9pt}
\begin{center}
\begin{tabular}{ | c | c | c | c | c  | c |} \hline
 shower & $\theta$ &$N_{e+\mu}$ & $c$ & $s_1$ & $s_2$  \\ \hline
\textbf{a}  & $30^\circ$ & 490538 & 0.000283 & -0.441 & -2.787   \\ \hline
\textbf{b}  & $45^\circ$ &1967956 & 0.000205 & -0.355 & -2.638   \\ \hline
\textbf{c}  & $52.5^\circ$ & 5834026 & 0.0000903 & -0.309 & -2.253   \\ \hline
\end{tabular}
\caption{The zenith angle, the shower size and the fit values for $c$, $s_1$ and $s_2$ for the combined density of the showers \textbf{a}, \textbf{b} and \textbf{c} as given in the text.}
\label{elmuparam}
\end{center}
\end{table}
\\
To obtain the LDF-A of the three showers we multiply the polar averaged LDF with $\cos \theta$ and replace $r$ for $k$. The final prediction for the LDF-A including the shift is
\begin{equation}\label{68}
\nu_{e+\mu}=N_{e+\mu} \cdot \cos \theta \cdot c \cdot \left( \frac{k}{30} \right)^{s_1} \left( 1+ \frac{k}{30} \right)^{s_2} \left( 1+\frac{k}{340} \right) \ ,
\end{equation}
where $k$ is as given by (\ref{61}) and with the parameter values as given in Table \ref{elmuparam}. 
\\ \\
For the three simulated showers we inspected the polar variation of the combined density at different radii. To this end the density was binned with bin size $\pi r / 12$ in the angular direction. For the bin size in the radial direction we took 1, 1, 2, 3, 4 and 5 m for radii 10, 20, 50, 100, 200 and 500 m respectively. The bins were just large enough to balance out to some extent the Poisson fluctuations in the density. For shower \textbf{a}, \textbf{b} and \textbf{c} the result is plotted in Fig. \ref{polar1630} through \ref{polar1852} together with the LDF-A prediction.
\begin{figure}[htbp]
\begin{center}
\includegraphics[width=12cm]{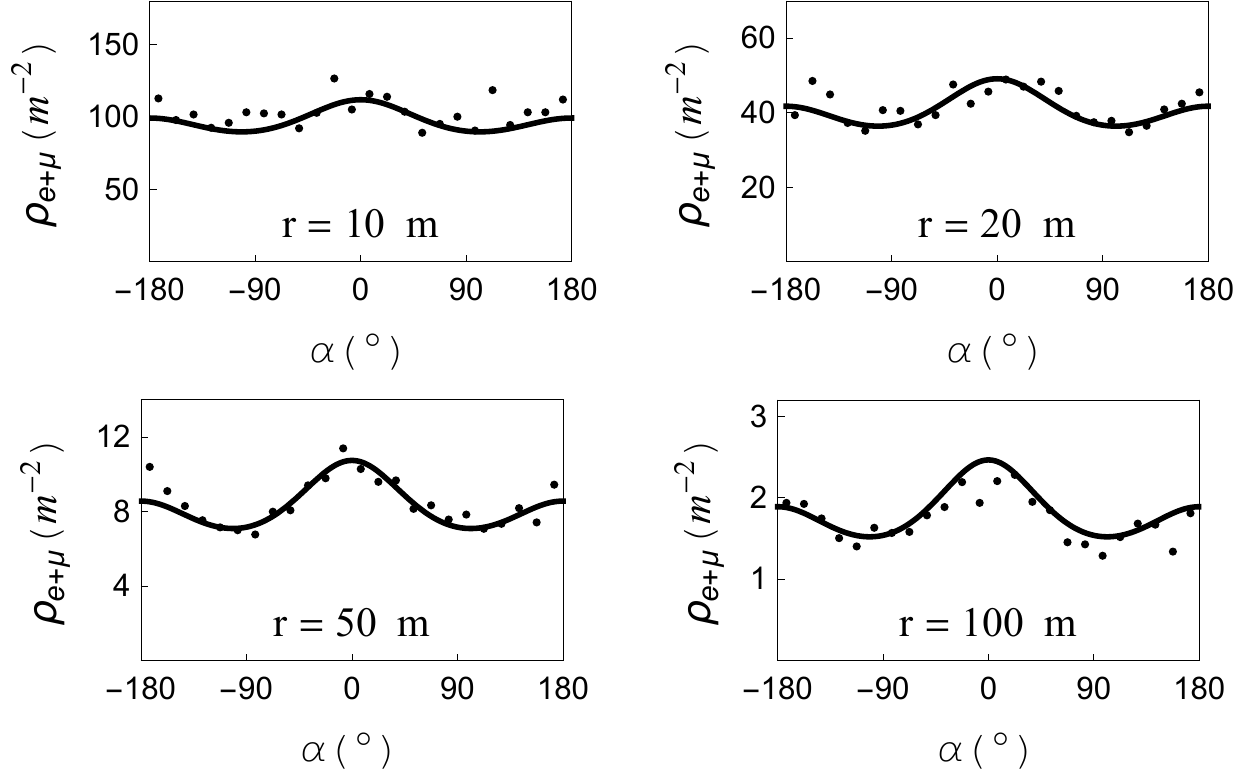}
\caption{The binned combined density of shower \textbf{a} (10$^{16}$ eV , zenith angle $30^\circ$) against polar angle for different radii (dots) and the constructed LDF-A prediction (solid).}
\label{polar1630}
\end{center}
\end{figure}
\begin{figure}[htbp]
\begin{center}
\includegraphics[width=12cm]{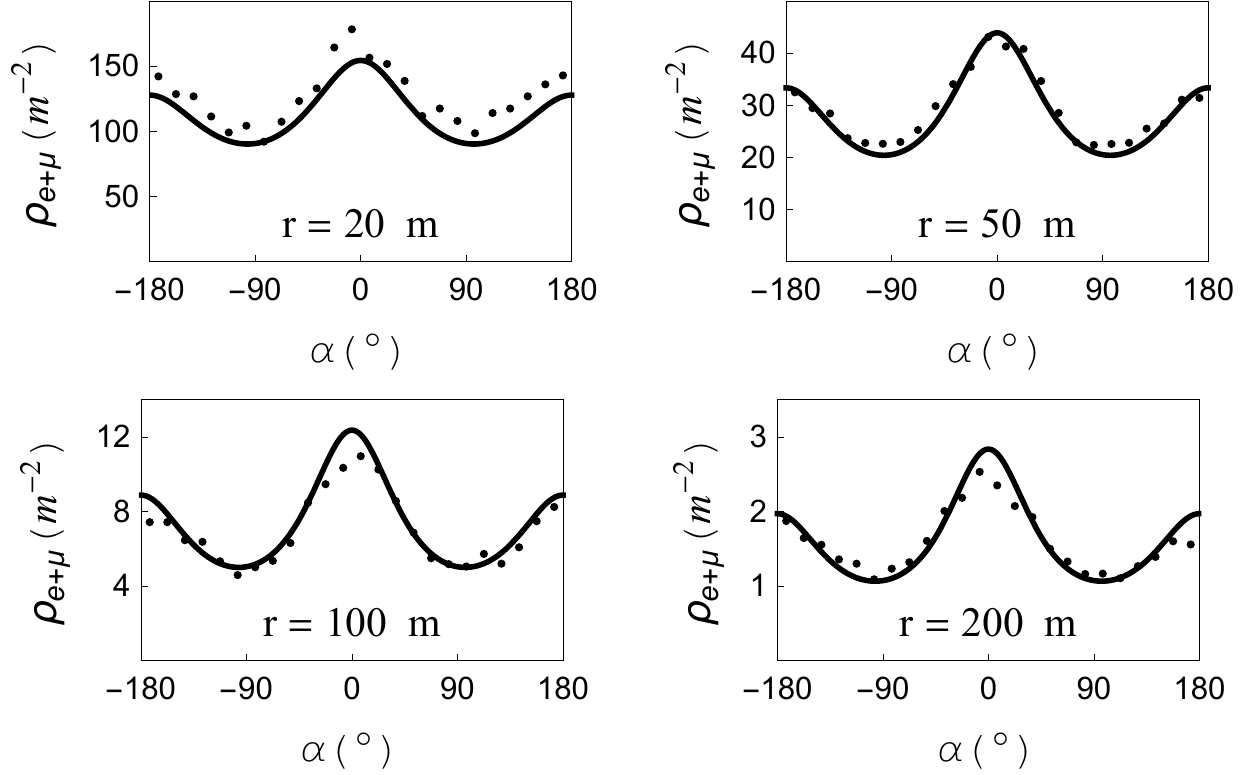}
\caption{The binned combined density of shower \textbf{b} (10$^{17}$ eV , zenith angle $45^\circ$) against polar angle for different radii (dots) and the constructed LDF-A prediction (solid).}
\label{polar1745}
\end{center}
\end{figure}
\begin{figure}[htbp]
\begin{center}
\includegraphics[width=12cm]{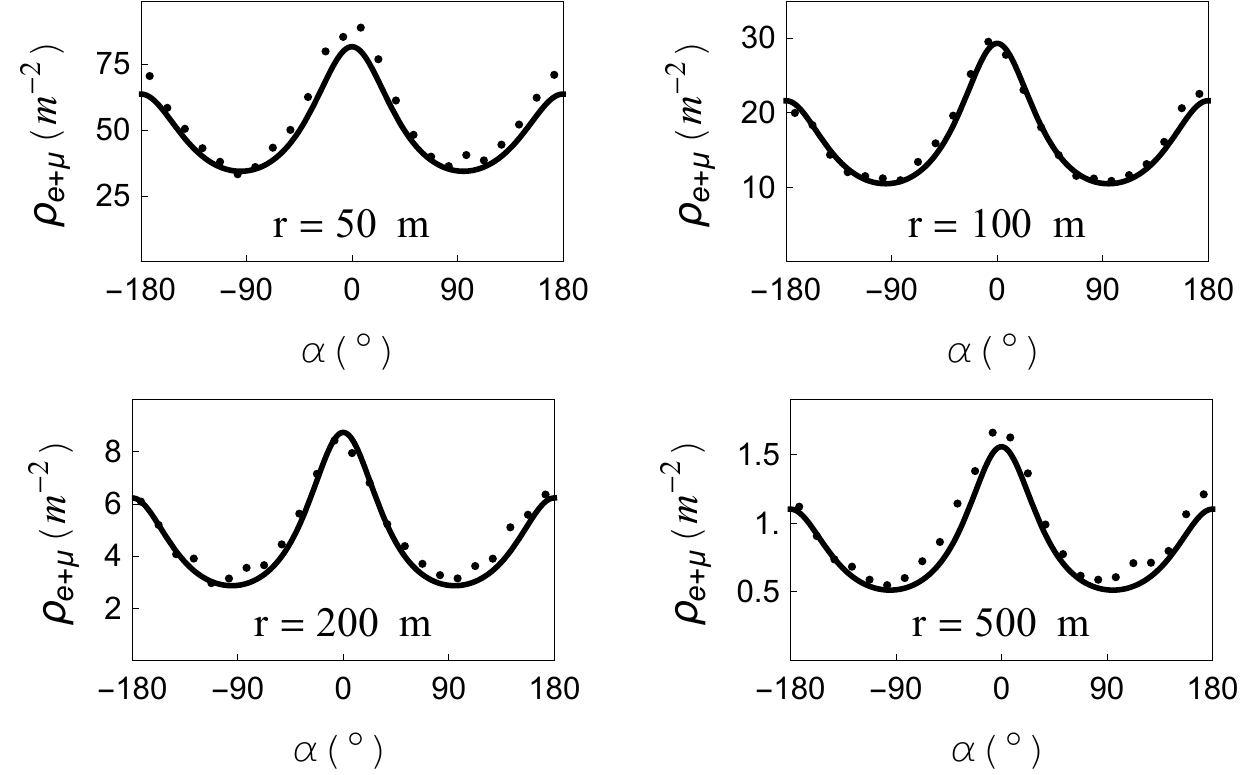}
\caption{The binned combined density of shower \textbf{c} (10$^{18}$ eV , zenith angle $52.5^\circ$) against polar angle for different radii (dots) and the constructed LDF-A prediction (solid).}
\label{polar1852}
\end{center}
\end{figure}
\\
We see the constructed LDF-A nicely follows the polar density. Angular independent deviations, such as in the upper left panel of Fig. \ref{polar1745}, are caused by the inaccuracy of $\rho$. Any inaccuracy in the underlying LDF will be reflected in the LDF-A. This does not take away that an LDF-A including the shift still follows the polar variation of the density better than an LDF-A without the shift.
\\ \\
The accuracy of the LDF-A for the three example showers suggests that it is probably sufficient to consider a single linear relation between $x_M$ and $k$ independent of zenith angle. If a better accuracy is desired one has to apply the power law (\ref{43}) with fit values for $A$ and $B$ as given in Table \ref{abparam}. The latter approach requires the fit coefficients either to be tabulated for different $\theta$ or to be parameterized to $\theta$ by means of a suitable function. Next to this, for a shift as given by the power law the equation \ref{59} would have been as follows
\begin{equation}\label{69}
 r^2 \cos^2 \alpha \cos^2 \theta - 2 r \cos \alpha \cdot A \cdot k^B  \sin \theta  + r^2 \sin^2  \alpha =k^2 \ .
\end{equation}
For $B$ a non-integer the latter equation has to be solved numerically for $k$.

\section{Summary}
For electron densities the shift of the center of elliptic iso-density contours is modeled. For combined densities of simulated showers the shift is determined for different zenith angles. An approximate linear relation between the shift and $k$ allows for an analytical solution for the asymmetric lateral density. The conversion of an LDF to an LDF-A including the shift is described. The conversion consists in two steps corresponding to the effects of the projection and of the attenuation. The first step is the multiplication of the LDF by $\cos \theta$, where $\theta$ is the zenith angle. The second step is to replace $r$ by $k$, where $k$ is given by Eq. (\ref{61}). The result is an LDF-A for the situation where the azimuth angle is equal to zero. The LDF-A for a non-zero azimuth angle $\phi$ requires a third, trivial step: replacing the polar angle $\alpha$ by $\alpha - \phi$.   
\\ \\
Zenith angle $60^\circ$ is a sort of transition point. Below this point there is a relatively large shift while the effect of the geomagnetic field is negligible.  Above the transition point the muon component becomes dominant. As a consequence the relative shift is smaller. At the same time the influence of the geomagnetic field rapidly grows with zenith angle. Above the transition point the consequences of the shift for the asymmetry of the density will be overwhelmed by the effect of the geomagnetic field. The influence of the geomagnetic field on very inclined air showers requires a different modeling \cite{valino2010, rodriguez2008, ave2000, Dembinski2010}.
\\ \\
The aim of the paper was to consider the situation for zenith angles smaller than $60^\circ$, where the shift is mainly governed by the attenuation of the electron component and therefore to a certain extend substantial. The inclusion of the shift leads to a more accurate description of an asymmetric polar density. It therefore may be worthwhile to take the shift into account for reconstruction purposes. It should be emphasized that the Eq. (\ref{44}) and the shift part of Eq. (\ref{61}) were derived for observation at sea level. For an observation level at a different altitude, the shift will be different. Even if the inclusion of the shift improves the accuracy of reconstruction only marginally, the contents of the paper may still contribute to the description and understanding of horizontal polar densities of inclined cosmic air showers. 

\section{Acknowledgements}
I wish to thank Dr. J.J.M. Steijger for his comments on an earlier draft of this paper. I am grateful to A. P. L. S. de Laat for his efforts in shower simulations. The work is supported by a grant from NWO (Netherlands Organization for Scientific Research).

\bibliography{References}

\begin{thebibliography}{widest-label}
\bibitem{Dova1999}
M.T. Dova, L.N. Epele and A. Mariazzi, Proc. 26\textsuperscript{th} ICRC, Vol. 1, 478 (1999).
\bibitem{Pryke}
C. Pryke, \textit{Asymmetry of Air Shower at Ground Level}, Auger technical note GAP-98-034, (1998).
\bibitem{Sima2011}
O. Sima et al., Nucl. Instr. and Meth. A \textbf{638}, 147 (2011).
\bibitem{Dova2003}
M.T. Dova, L.N. Epele and A. Mariazzi, Astropart. Phys. \textbf{18}, 351 (2003).
\bibitem{CiampaClay}
D. Ciampa and R.W. Clay, J. Phys. G: Nucl. Phys. \textbf{14} 787 (1988).
\bibitem{Antoni2003}
T. Antoni et al., Astropart. Phys. \textbf{19}, 703 (2003).
\bibitem{valino2010}
I. Valino et al., Astropart. phys. \textbf{32}, 304 (2010).
\bibitem{corsika1}
D. Heck et al. , Wissenschaftliche Berichte, Forschungszentrum Karlsruhe FZKA 6019 (1998).
\bibitem{ostap1}
S. Ostapchenko, Nucl. Phys. Proc. Suppl. \textbf{151}, 143 (2006).
\bibitem{ostap2}
S. Ostapchenko, Prog. Theor. Phys. Suppl. \textbf{193}, 204 (2012).
\bibitem{gheisha1}
H. Fesefeldt, Report PITHA-85/02, RWTH Aachen, 1985.
\bibitem{garcia-pinto2009}
D. Garc\'ia-Pinto et al., Proc. 31\textsuperscript{st} ICRC, \L\'OD\'Z (2009).
\bibitem{apcworkshop}
T. Pierog et al., \textit{(Future of) Shower Physics}, 5\textsuperscript{th} Workshop on Air Shower Detection at High Altitude, APC, Paris (2014).
\bibitem{cillis}
A. Cillis and S.J. Sciutto, J. Phys. G: Nucl. Part. Phys. \textbf{26} 309 (2000).
\bibitem{K&N}
K. Kamata, J. Nishimura, Suppl. Prog. Theor. Phys. \textbf{6}, 93 (1958).
\bibitem{Greisen}
K. Greisen,  Prog. Cosmic Ray Phys. \textbf{3}, 3 (1956).
\bibitem{N&W}
M. Nagano, A.A. Watson, Rev. Mod. Phys. \textbf{72}, 689 (2000).
\bibitem{Apel}
W.D. Apel et al., Astropart. Phys. \textbf{24}, 467 (2006).
\bibitem{Vernov1968}
S.N. Vernov et al., Can. J. Phys. \textbf{46}, s197 (1968).
\bibitem{Greisen1960}
K. Greisen, Ann. Rev. Nucl. Sci. \textbf{10}, 63 (1960).
\bibitem{Antoni2001}
T. Antoni et al., Astropart. Phys. \textbf{14}, 245 (2001).
\bibitem{greisen1960}
K. Greisen, Ann. Rev. Nucl. Sci. \textbf{10}, 63 (1960).
\bibitem{rodriguez2008}
G. Rodriguez, J. Phys.: Conf. Ser. \textbf{116} 012006 (2008).
\bibitem{ave2000}
M. Ave, R.A. V\'azquez and E. Zas, Astropart. Phys. \textbf{14} 91 (2000).
\bibitem{Dembinski2010}
H.P. Dembinski et al., Astropart. Phys. \textbf{34} 128 (2010).


\end{thebibliography}

\end{document}